\def\N{\hbox{I \kern-.5em N}}
\def\M{\hbox{I \kern-.5em M}}
\def\J{\hbox{I \kern-.5em J}}
\def\B{\hbox{I \kern-.5em B}}
\def\Q{\hbox{I \kern-.5em Q}}
\def\D{\hbox{I \kern-.5em D}}
\def\G{\hbox{I \kern-.5em G}}
\def\E{\hbox{I \kern-.5em E}}
\def\F{\hbox{I \kern-.5em F}}
\def\A{\hbox{I \kern-.5em A}}
\def\H{\hbox{I \kern-.5em H}}
\def\F{\hbox{I \kern-.5em F}}
\def\L{\hbox{I \kern-.5em L}}
\def\P{\hbox{I \kern-.5em P}}
\def\X{\hbox{I \kern-.5em X}}
\def\R{\hbox{I \kern-.5em R}}

\documentclass[twocolumn]{autart}  

\usepackage{pgfplots,enumitem,pgf,cite,tikz,graphics,graphicx,color,dsfont,amsmath,epsfig,epstopdf,float,amssymb,bm,mathrsfs,upgreek,algorithm,algpseudocode,pifont,soul,adjustbox} 
\usepackage[utf8]{inputenc}
\usepackage{pgfplots} 
\usepackage{mathtools}
\usepackage{pgfgantt}
\usepackage{pdflscape}
\usepackage{bbm}
\usepackage{mathbbol}
\pgfplotsset{compat=newest} 
\pgfplotsset{plot coordinates/math parser=false}
\newlength\fwidth
\usepackage{lipsum}
\usepackage{ntheorem}
\usepackage{balance}

\graphicspath{{./Figures/}}
\usepackage[font=footnotesize, caption=false]{subfig}
\DeclareMathAlphabet{\mathpzc}{OT1}{pzc}{m}{it}
\newtheorem{remark}{Remark}

\newtheorem{definition}{Definition}
\newtheorem{theorem}{Theorem}
\newtheorem{corollary}{Corollary}

\newtheorem*{proof-non}{Proof}
\newtheorem{lemma}{Lemma}
\usetikzlibrary{positioning,arrows,automata,backgrounds,fit}
\pgfdeclarelayer{background}
\pgfdeclarelayer{foreground}
\pgfsetlayers{background,main,foreground}
\usepackage{lipsum}

\newcommand\blfootnote[1]{%
  \begingroup
  \renewcommand\thefootnote{}\footnote{#1}%
  \addtocounter{footnote}{-1}%
  \endgroup
}

\definecolor{Green}{rgb}{0,0.5,0}

 \usepackage[natbibapa]{apacite}
\begin{document}
\setlength{\belowdisplayskip}{1pt} \setlength{\belowdisplayshortskip}{1pt}
\setlength{\abovedisplayskip}{1pt} \setlength{\abovedisplayshortskip}{1pt}
\begin{frontmatter}
  
\title{Nonlinear Observability via Koopman Analysis: \\
Characterizing the Role of Symmetry} 

\author[JZF]{Afshin Mesbahi}, \ead{amesbahi@uw.edu}                                     
\author[JZF]{Jingjing Bu}, \ead{bujing@uw.edu}             
\author[JZF]{Mehran Mesbahi}\ead{mesbahi@uw.edu}           

\address[JZF]{William E. Boeing Department of Aeronautics and Astronautics, University of Washington,WA 98195}  

	\begin{keyword}                           
		Nonlinear observability; Koopman operator; Symmetry          
	\end{keyword}                             

\begin{abstract}         	
This paper considers the observability of nonlinear systems from a Koopman operator theoretic
perspective--and in particular--the effect of symmetry on observability.
We first examine an infinite-dimensional linear system (constructed using independent Koopman eigenfunctions) such that its observability is equivalent to the observability of the original nonlinear system. 
Next, we derive an analytic relation between symmetry and nonlinear observability;
it is shown that symmetry in the nonlinear dynamics is reflected
in the symmetry of the corresponding Koopman eigenfunctions,
%
as well as presence of repeated Koopman eigenvalues.
We then proceed to show that the loss of observability in symmetric nonlinear systems can
be traced back to the presence of these repeated eigenvalues.
%
%
In the case where we have a sufficient number of measurements, 
the nonlinear system remains unobservable when these functions have symmetries that mirror those of the dynamics.
The proposed observability framework provides insights into the minimum number of the measurements 
needed to make an unobservable nonlinear system, observable.
The proposed results are then applied to a network of nano-electromechanical oscillators coupled via a symmetric interaction topology.
\end{abstract}
\end{frontmatter}


\section{Introduction} \label{Section_Introduction}
\blfootnote{A preliminary version of this work has appeared in the 2019 American Control Conference~\cite{Mesbahi_ACC_2019}.}
Dynamic systems are described by a set of interacting internal variables,
collectively referred to as the system state.
The interdependence between internal variables, in turn, provides the possibility of reconstructing the state
by tracking only a subset of these variables.
%
A natural question that arises in this context pertains to the (minimal) number of measurements 
required to allow estimating the entire (internal) state.
The observability problem addresses this issue by establishing connections
between the state dynamics and measurements in order to uniquely deduce 
the state (or its initial condition).

For linear systems, observability is examined via necessary and sufficient linear algebraic conditions--{each method providing unique insights supported by efficient algorithmic realizations}--in order to determine if measurements are adequate for such a reconstruction.
Analogously, nonlinear observability can be examined via theoretical or numerical methods.
Theoretical observability analysis utilizes constructs from algebra and differential geometry.
Most of the existing approaches in this direction provide sufficient conditions  for observability by computing the dimension of the subspace spanned by the gradients of the { Lie derivatives} of the measurements \cite{Hermann_TAC_1977,Martinelli_TAC_2018,Zabczyk_Book_2007,Bartosiewicz_Automatica_2016,Kawano_Automatica_2017}.
Differential geometric approaches to nonlinear observability, however, are generally difficult to realize
in terms of efficient algorithms, {nor are they amenable to online, data-driven scenarios.}
The ``empirical" observability provides an alternate framework to examine nonlinear observability,
leading to a numerical procedure for computing the rank of the observability Gramian around a nominal trajectory of a nonlinear system \cite{Aston_IJBC_1995,Lall_IJRNC_2002,Krener_CDC_2009,Powel_CDC_2015}.
In the meantime, empirical observability might not be applicable in certain scenarios,
as it requires the ability to simulate
the system from perturbed initial conditions for each state, and comparing the corresponding measurements.

This work delves into characterizing the effect of {discrete} symmetries on nonlinear observability.
{Symmetry is a fundamental property of many natural and technological systems~\cite{Field_Book_2009,Golubitsky_Book_2003}.
For example, symmetries are common in systems such as social, cellular, and
oscillatory networks \cite{Emenheiser_Chaos_2016,Alaeddini_CDC_2018}.
For linear systems, fundamental links between {discrete} symmetries (parametrized in terms of an automorphism group)
and observability have been studied~\cite{Rahmani_SIAM_2009,Chapman_CDC_2014,Chapman_CDC_2015}.
Characterizing the effect of {discrete} symmetries on nonlinear observability, in the meantime,
requires more intricate analysis.

It is known that nonlinear systems with {discrete symmetries} may become unobservable~\cite{Liu_PNAS_2013,Letellier_Chaos_2002,Martinelli_TRO_2011}.
In the case that a control-affine nonlinear system is symmetric, the unobservable subspace can in fact be identified using the kernel of the observability matrix \cite{Martinelli_TRO_2011}.
Through numerical simulations, it has also been shown that the existence {of} {discrete} symmetries in a network of nonlinear systems may decrease its observability and controllability \cite{Whalen_PhysRevX_2015}.
Of special interest in this work are observations reported in the literature that the so-called ``reflectional" symmetries in the network may lead to unobservability, and networks with only ``rotational" symmetries may remain observable.
One of the main objectives of this paper is to theoretically explain why rotational and reflectional symmetries have different effects on the observability of nonlinear systems.

The approach adopted in this work for understanding connections between symmetries and observability
is based on the Koopman operator formalism {initiated in \cite{Mezic_PD_2004,Mezic_ND_2005}, utilizing the operator theoretic representation of nonlinear dynamics introduced in \cite{Koopman_PNAS_1931}}.
Over the past decade, there has been tremendous interest to utilize spectral properties of the Koopman operator for understanding complex nonlinear phenomena~\cite{Mathias_CDC_2017,Mesbahi_ACC2_2019,Bollt_SIAM_2018,Mauroy_TAC_2020,Sootla_TAC_2017,Mauroy_TAC_2016,Korda_Automatica_2018,Sootla_TAC_2018,Sootla_Automatica_2018,Surana_CDC_2016,Kaiser_CDC_2018,Proctor_SIAM_2018,Mauroy_SIAM_2017}. 
The Koopman operator encodes the time evolution of the observable functions along the trajectories of the nonlinear system
\cite{Schmid_JFM_2010,Mezic_ARFM_2013,Tu_JCD_2014}; furthermore, this operator can be
approximated using data-driven techniques \cite{Williams_JNS_2015,Williams_JCD_2015,Lusch_Nature_2018,Li_Chaos_2017}.
Consequently, the Koopman operator framework has become an attractive technique for analyzing dynamical systems using the time-series data.
This approach has also been used in system-theoretic settings. For example,
controllability and observability of nonlinear systems with affine structure have been 
studied based on a truncation or approximation of the infinite-dimensional linear system in the Koopman space \cite{Surana_CDC_2016,Goswami_CDC_2017}.

In this paper, we first examine the representation of
the nonlinear system as an infinite-dimensional linear system using independent Koopman eigenfunctions.
Although, this representation is infinite-dimensional, there exist necessary and sufficient conditions 
for checking its observability due to its linearity.
These conditions can be checked through the rank of the so-called observability matrix~\cite{Triggiani_SIAM_1976,Klamka_Book_1991,Son_JMAA_1990,Curtain_Book_2012}.
Subsequently, we show that analyzing the observability of the original nonlinear system is equivalent to checking the observability of this ``transformed" infinite-dimensional linear system.

Analyzing the observability problem from the Koopman operator perspective provides a ``spectral"
bridge between {discrete} symmetries and observability of nonlinear systems.
{This is accomplished by showing how symmetries in the dynamics 
are reflected in the spectra of the Koopman operator~\cite{Mesbahi_ACC_2019}.}
%
%
%
%
Using this spectral approach, one can also uncover the distinct
effects of  the so-called rotational and reflectional symmetries on nonlinear observability, 
previously reported using simulation studies in~\cite{Whalen_PhysRevX_2015}.
%
%
For example, it is shown that Koopman eigenfunctions with reflectional symmetries lead to 
repeated Koopman eigenvalues in contrast with those with only rotational symmetries.
{In the case of reflectional symmetries, we show that the loss of observability can then be traced back
to the presence of repeated eigenvalues.}
In this case, the number and structure of measurements are the main indicators 
for evaluating observability.
In particular, if the measurements mirror the same type of symmetry as the underlying dynamics, repeated Koopman eigenvalues lead to trajectories from distinct initial conditions that are indistinguishable from each other through the respective measurements.
However, nonlinear systems with repeated Koopman eigenvalues can still be observable if the measurements do not have the same type of symmetries as the underlying dynamics.
In this case, the number of repeated Koopman eigenvalues is critical to determine the minimum number of 
measurements required to ``observe" the internal state.
In the meantime, rotational symmetries do not lead to repeated Koopman eigenvalues.
Therefore, nonlinear systems with only rotationally symmetric Koopman eigenfunctions may still be observable.

The remainder of the paper is organized as follows.
\S \ref{Section_Math} contains mathematical preliminaries on the observability problem and the Koopman operator.
In \S \ref{Subsection_Infinite_Dim_Sys}, we provide a procedure to transform the nonlinear system into an infinite-dimensional linear system by using independent Koopman eigenfunctions.
In \S \ref{Section_Observability}, we formulate the nonlinear observability problem 
in terms of the observability of a 	``transformed" infinite-dimensional linear system.
\S \ref{Section_Symmetry_Koopman} presents our results on the connection between {discrete} symmetries in 
the nonlinear system and its observability via the spectral properties of the
corresponding Koopman operator.
In \S \ref{Section_Examples}, we provide three examples to demonstrate the applicability
of our theoretical results. One of these examples pertain to the observability 
analysis of coupled nanoelectromechanical systems (NEMS) on a ring topology.
Finally, \S \ref{Section_Conclusion} includes the concluding remarks and future directions 
of this work.


\section{Preliminaries} \label{Section_Math}
{In this section, we provide an overview of the notation and preliminary background on the observability analysis and the Koopman operator.}

Let $\mathbb{N}$, $\mathbb{R}$, and $\mathbb{C}$ denote the natural, real, and complex numbers, respectively; $\mathbb{j}=\sqrt{-1}$ and real and imaginary parts of a complex number
will be denoted by $\mbox{\bf Re}$ and $\mbox{\bf Im}$, respectively.
{$A \setminus B$ denotes the relative complement of the set $B$ with 
respect to set $A$, the set of elements that are members of $A$ but not $B$.}
The notation $|c|$ and $\angle c$ refer to the magnitude and phase of the complex number $c$.
The inner product between a pair of vectors $\bm{x}$ and $\bm{y}$ is {denoted} by $\langle \bm{x} , \bm{y} \rangle$.
The identity matrix is denoted by $\bm{I}$ (its dimension implied in the context); {$\mathbbm{1}$ denotes the vector of all ones.;}
$\mbox{\bf rank}(\bm{A})$ and $\bm{A}^\top$ 
represent the rank of the matrix ${\bm{A}}$ and its
transpose, respectively.
The Lie derivative of a tensor field, e.g., a scalar function
$\psi$, along the vector field $f$, is denoted by $\mathcal{L}_f \psi = \big \langle \nabla \psi,  f \big \rangle$.

\subsection{Observability}
{
We consider a class of nonlinear systems of the form,
\begin{equation}\label{eq_auto_system}
\begin{split}
 \dot{\bm{x}}(t) = f \big( \bm{x}(t) \big) , \quad \bm{y} (t) = h \big( \bm{x}(t) \big),
\end{split}
\end{equation}
where the state space $\mathcal{M} \subseteq \mathbb R^n$ is a smooth manifold of dimension $n$, $\bm{x}$ denotes the system state evolving
in $\mathcal{M}$, $f: \mathcal{M} \rightarrow \mathcal{M}$ is %
{the $C^{\infty}$-smooth vector field}%
, and $h: \mathcal{M} \rightarrow \mathbb{R}^q$ is a set of $C^{\infty}$-smooth (nonlinear) measurements consisting of $q$ scalar-valued functions.}
%
Given $t \geq 0$, the flow map $\Phi_t \big(\bm{x}_{0} \big): \mathcal{M} \rightarrow \mathcal{M}$ evolves the state $\bm{x}_{0}=\bm{x} (t_0)$ to the future state $\bm{x} (t+t_0)$ as,
\begin{equation}\label{eq_flow_map}
\Phi_t \big(\bm{x}(t_0) \big) = \bm{x} (t+t_0)  =  \bm{x} (t_0) + \int_{t_0}^{t+t_0} f \big( \bm{x}(\tau) \big) d\tau .
\end{equation}
The flow map, in conjunction with the measurements, induce a composition map
$h \circ \Phi_t$ that is of particular interest in the context of observability analysis.
In fact, observability of (\ref{eq_auto_system}) pertains to the ability of 
identifying the initial state ${\bm x}_0$ from the image of the composition map $h \circ \Phi_t$ 
for some $t>0$.\footnote{That is, in principle, there is an algorithm that can identify the initial condition from the observation time history.}
Although, various notions of observability for linear systems (consisting of linear state dynamics, augmented with linear measurements) are equivalent and can be tested using for example, the rank of the observability matrix,
there are several distinct notions of observability for nonlinear systems.
In this work, we use the notion of nonlinear observability %
{as} %
 adopted in~\cite{Hermann_TAC_1977}.
 \vspace{.05in}
\begin{definition}
{The system (\ref{eq_auto_system}) is locally weakly observable {at} $\bm{x}_0$ 
if there exists a neighborhood $\mathcal{D}$ containing $\bm{x}_0$ such that for every open neighborhood $\mathcal{U}$ of  $\bm{x}_0$ contained in $\mathcal{D}$ and for every state $\bm{x}_1 \in \mathcal{U}$ ($\bm{x}_0 \neq \bm{x}_1$),
\begin{equation*}\label{eq_definition_obs}
h \circ \Phi_t \big(\bm{x}_0 \big) \neq h \circ \Phi_t \big(\bm{x}_1 \big),
\end{equation*}
for some (finite) $t > 0$.
The system (\ref{eq_auto_system}) is called locally weakly observable if it is locally observable for all $\bm{x}_0 \in \mathcal{M}$.
The system (\ref{eq_auto_system}) is called observable if it is locally weakly observable and the corresponding neighborhoods can be taken as $\mathcal{M}$.}
\end{definition}
Observability as defined above
essentially dictates that distinct initial conditions should lead to distinct measurement trajectories 
for some $t>0$.
{The standard approach to address nonlinear observability utilizes constructs from differential geometry.
This is in view of the fact that the observability of the system (\ref{eq_auto_system}) can be expressed based on the measurement $h$ and its higher-order Lie derivatives with respect to the differential flow map $f$, or equivalently, based on the span of time derivatives of the measurement $h$ along all possible trajectories \cite{Hermann_TAC_1977,Zabczyk_Book_2007}.
The higher order Lie derivatives of $\psi$ with respect to the vector field $f$ are defined as $\mathcal{L}^k_f \psi = \big \langle \nabla \mathcal{L}^{k-1}_f  \psi,  f \big \rangle$, where $\mathcal{L}^0_f \psi = \psi$ and $k \in \mathbb{N}$.
{Then system (\ref{eq_auto_system}) is locally weakly observable at $\bm{x}_0$ if
$$
\text{\bf rank}\;
\bigg(
\frac{d}{d \bm{x}}
\begin{bmatrix}
\mathcal{L}^{0}_f h \\
\vdots \\
\mathcal{L}^{k-1}_f h 
\end{bmatrix}_{|\bm{x} = \bm{x}_0}
\bigg) 
= n;
$$
here, $k$ represents the number of required higher order Lie derivatives
for determining the observability by checking the rank condition of the corresponding
$kq \times n$-dimensional matrix \cite{Hermann_TAC_1977,Zabczyk_Book_2007}}.
The minimum number of Lie derivatives required for satisfying the rank condition is $k=n$.

In the meantime numerical approaches for testing nonlinear observability, via for example, the empirical observability Gramian, have become popular in practice \cite{Aston_IJBC_1995,Lall_IJRNC_2002,Krener_CDC_2009,Powel_CDC_2015}.
In this direction, let us consider perturbations of the state $\bm{x}_0$ assuming the form $\bm{x}_{\pm i} = \bm{x}_0 \pm \epsilon \bm{e}_i$, and the corresponding measurements as $\bm{y}_{\pm i}$, where $\epsilon$ is a real positive scalar and $\bm{e}_i \in \mathbb{R}^n$ denotes the unit vector with one at the $i$th entry  and zero elsewhere.
The empirical observability Gramian at $\bm{x}_0$ is then the $n \times n$-dimensional matrix, $$ \bm{G}_t^\epsilon  \big(\bm{x}(t_0) \big) = \frac{1}{4 \epsilon^2}  \int_{0}^{t} \Phi_\tau^\epsilon  \big(\bm{x}(t_0) \big)^\top \Phi_\tau^\epsilon  \big(\bm{x}(t_0) \big) d \tau,$$ where $$ \Phi_\tau^\epsilon = \begin{bmatrix}
\bm{y}_{+1} - \bm{y}_{-1} ,& \ldots, & \bm{y}_{+n} - \bm{y}_{-n}
\end{bmatrix}.$$
{It is known that the system (\ref{eq_auto_system}) is locally weakly observable at $\bm{x}_0$} if $$\text{\bf rank} \Big( \lim_{\epsilon \to 0} \bm{G}_t^\epsilon  \big(\bm{x}(t_0) \big) \Big)  =n.$$
We note that empirical observability analysis can be expensive as it requires computing solutions to a nonlinear system from $2n$ distinct initial conditions; in some scenarios, perturbing the initial condition at will is also prohibitive.

In this work, we examine the observability of nonlinear systems from the perspective of the Koopman operator,  a linear operator that facilitates a {\em spectral approach} for understanding nonlinear phenomena. One of the advantages of adopting an operator theoretic approach for nonlinear observability analysis is the ability to differentiate unobservable and observable dynamics based on the corresponding spectral decomposition.
Moreover, this approach allows for a modal perspective on how the existence of {discrete} symmetries leads to unobservability of the system,
analogous to similar results for linear systems such as {Popov-Belevitch-Hautus test~\cite{Zabczyk_Book_2007}.}
Lastly, the Koopman analysis facilitates reasoning about 
the minimum number of measurements needed to make an otherwise unobservable system, observable.

\subsection{Koopman Operator} \label{Subsection_Koopman}
{We use the terminology of an ``observation" function $\phi: \mathcal{M}  \rightarrow  \mathbb{C}$ as a scalar-valued $C^{1}$ function on the state space;\footnote{Note that in our terminology, measurements and observation functions refer to rather distinct objects; 
{The measurement is an output function as typically used in dynamics and control, physically realized for example by a sensor.  An observation function--on the other hand--is defined on the state space of the dynamical system for the purpose of a Koopman analysis~\cite{Williams_JNS_2015,Mauroy_TAC_2016}.} } in turn, $\mathcal F$ denotes the collection of all such observation functions.}
For a given $\phi \in \mathcal{F}$ and nonlinear flow map $\Phi_t$ (\ref{eq_flow_map}), the Koopman operator $\mathcal{K}_t: \mathcal{F}  \rightarrow \mathcal{F}$ is defined as the map for which, $$ \mathcal{K}_t \phi \big( \bm{x} \big) = \phi  \circ \Phi_t \big(\bm{x}\big),$$
for every $\bm{x} \in \mathcal{M}$ {\cite{budivsic2012applied,Mezic_ARFM_2013} }; see Figure~\ref{koopman-diagram}.
For a given flow map $\Phi_t$ and the observation function $\phi$, we denote the corresponding
Koopman operator as $\mathcal{K}^{\Phi,\phi}_t$;
in the case that both $\Phi$ and $\phi$ are clear from the context, we simply write $\mathcal{K}_t$ to
denote the corresponding Koopman operator.
%
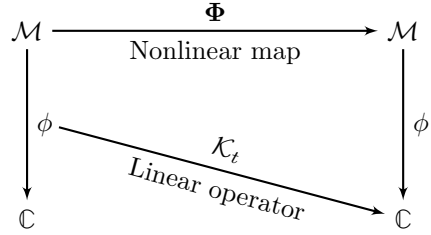
\begin{figure}
\begin{center}
\begin{tikzpicture}[>=latex',node distance = 2.5cm]
\node (M2) {$\mathcal{M}$};
\node [left of = M2, node distance = 5cm] (M1) {$\mathcal{M}$};
\node [below of = M1, node distance = 2.5cm] (K1) {$\mathbb{C}$};
\node [below of = M1, node distance = 1.25cm] (empty) {};
\node [right of = empty, node distance = 0.3cm] (M1K1) {};

\draw[->,thick] (M1) to node[right] {$\phi$} (K1);

\node [below of = M2, node distance = 2.5cm] (K2) {$\mathbb{C}$};
\draw[->,thick] (M2) to node[right] {$\phi$} (K2);

\draw [->,thick] (M1) to node[above] {$\bm{\Phi}$} node[below] {Nonlinear map} (M2);
\draw [->,thick] (M1K1) to node[above,sloped] {$\mathcal{K}_t$}  node[below,sloped] {Linear operator} (K2);
\end{tikzpicture}
\caption{The Koopman operator evolves the observable of a nonlinear system.} \label{koopman-diagram}
\end{center}
\end{figure}
%


We first note that the Koopman operator associated with any (potentially nonlinear) flow map is linear; that is, the Koopman operator corresponding to the linear combination of two observation functions is the linear combination of the Koopman operator applied to each.
{Consistent with the definition of the Koopman operator $\mathcal{K}_{t}$,  the variable,
$$\bar{\phi}_t \big( \bm{x} \big) =  \mathcal{K}_{t} \phi \big( \bm{x} \big)$$ 
is the solution of the partial differential equation,
\begin{equation*}
\begin{split}
\frac{\partial \bar{\phi}_t \big(\bm{x} \big)}{\partial t} = \mathcal{L}_f \bar{\phi}_t \big(\bm{x} \big), \quad 
\bar{\phi}_0 \big(\bm{x} \big) = \phi \big(\bm{x}_0 \big),
\end{split}
\end{equation*}
where $\bm{x}_0$ is the initial condition of (\ref{eq_auto_system})~\cite{Lasota_Book_1994};
we refer to {$\mathcal{L}_f$} as the Koopman generator \cite{Mezic_ARFM_2013,budivsic2012applied}};
note that the Koopman generator satisfies the equation,
\begin{equation*}
\mathcal{L}_f \psi_i \big( \bm{x} \big) = \lambda_i \psi_i \big( \bm{x} \big), \quad i \in \mathbb{N},
\end{equation*}
where $\lambda_i \in \mathbb{C}$ and $\psi_i: \mathcal{M}  \rightarrow  \mathbb{C}$ are the Koopman eigenvalue and eigenfunction, respectively.
As such, the Koopman operator can be described by its spectral properties, namely its eigenfunctions and eigenvalues.
Let $(\lambda_i, \psi_i)$ be an eigenvalue/eigenvector pair for  $\mathcal{K}_{t}$;
then for all  $\bm{x} \in \mathcal{M}$, 
\begin{equation*}\label{eq_eigendecomposition}
\mathcal{K}_{t} \psi_i \big(  \bm{x}\big) =e^{\lambda_i t} \psi_i \big( \bm{x}\big).
\end{equation*}
Note that we are implicitly restricting our attention to the case where
the Koopman operator has a countable {point} spectra ($i \in \mathbb{N}$).
}

Even though the Koopman operator is linear, it is generally infinite-dimensional for nonlinear as well as linear systems.
As such, the Koopman operator may have point and continuous spectrum \cite{Mezic_ARFM_2013}.
{
It is known that the discrete spectrum characterizes the almost periodic part of dynamical systems, while the continuous part correspondeds to either a shear flow behavior or chaotic dynamics \cite{Govindarajan_ArXiv_2018,Mezic_ARFM_2013,Schmid_book_2001,Sharma_PRF_2016}.
In other words, 
the discrete spectrum describes the behavior of the dynamical system over isolated frequencies \cite{Lusch_Nature_2018}.
In this work, we only consider the point spectra of the Koopman operator in our observability analysis as it is sufficient 
for describing the evolution of observables in many physical and engineering systems \cite{Susuki_ITPS_2012,Mauroy_TAC_2016,Surana_CDC_2016,Goswami_CDC_2017,Sootla_Automatica_2018,Korda_Automatica_2018}.
}

Let $\bm{\Upsilon}: \mathcal{M}  \rightarrow  \mathbb{C}^{n_o}$ be an ${n_o}$-tuple of observation functions.
If this observation $\bm{\Upsilon}  \big( \bm{x} \big) $ lies within the closure of the span of Koopman eigenfunctions, the vector-valued observation can be expressed as,
 \begin{equation}\label{eq_obs}
\bm{\Upsilon}  \big( \bm{x} \big) =\sum_{j=1}^\infty \psi_j \big( \bm{x} \big) \bm{v}_j ,
\end{equation}
where $\bm{v}_j\in \mathbb{C}^{n_o}$ is a set of vector-valued coefficients {\cite{Mezic_ARFM_2013,budivsic2012applied}} {and the convergence is interpreted as point-wise absolutely convergent in the $2$-norm; namely, $\sum_{j=1}^{\infty} \|\psi_j \big(  {\bm x}\big) {\bm {v}_j}\|_2 < \infty$.}
Although Koopman eigenvalues and eigenfunctions are intrinsic characteristics of the dynamics (\ref{eq_auto_system}), vector-valued coefficients $\bm{v}_j $ depend on the choice of the observable $\bm{\Upsilon}$.
For the case of full-state observable $\bm{\Upsilon}  \big( \bm{x} \big) = \bm{x}$, the corresponding vector-valued coefficients $\bm{v}_j$'s are called the Koopman modes, that can be viewed as components of the projection of the state on the span of Koopman eigenfunctions.

We note that if the Koopman operator $\mathcal{K}_{t}$ has a pair of eigenvalues $\lambda_1$ and $\lambda_2$ with the corresponding eigenfunctions $\psi_1$ and $\psi_2$, then $\mathcal{K}_{t}$ also has eigenvalues $\alpha_1 \lambda_1+\alpha_2 \lambda_2$ and eigenfunctions $c \psi_1^{\alpha_1} \psi_2^{\alpha_2}$, where $c \in \mathbb{C}$ and $\alpha_1, \alpha_2 \in \mathbb{N}$ {\cite{Mezic_ARFM_2013,budivsic2012applied}}.
A set of independent Koopman eigenfunctions whose associated Koopman modes are nonzero will be referred to as the Koopman set $\bm{\Psi}$.\footnote{Note that a Koopman set is not necessary of finite cardinality.}
{An important assumption in our subsequent analysis is that {full-state} observation vector $\bm{\Upsilon}  \big( \bm{x} \big)=\bm{x} $ lies in the span of the Koopman set $\bm{\Psi}$.}
As such, nonlinear systems of interest in this work can be described by the eigenspace of the 
Koopman operator via the Koopman eigenfunctions.
To summarize, in this work the following statements hold by construction: 
(a) the Koopman {set} $\bm{\Psi}$ is nonempty, as the constant function $c \in \mathbb{C}$ is always a trivial eigenfunction of the Koopman operator with zero eigenvalue,
(b) when referring to Koopman eigenfunctions $\psi_1, \psi_2, \ldots \in \bm{\Psi}$, we are implicitly assuming their linear independence,
(c) Koopman modes $\bm{v}_1, \bm{v}_2, \ldots,$ associated with Koopman eigenfunctions $\psi_1, \psi_2, \ldots  \in \bm{\Psi},$ are nonzero; this follows from our adopted definition for the Koopman set.

Koopman operator facilitates the representation of a nonlinear system as 
an infinite-dimensional
linear system, once the set of observation functions is fixed. The setup has been used extensively
in model identification, particularly in the context of the so-called dynamic mode decomposition (DMD), where a finite-dimensional representation of the Koopman operator is constructed using the time-series data.
It is thus of interest to explore the extent by which the Koopman operator representation of a
nonlinear system can facilitate their system-theoretic analysis.
{Such an approach has been explored in~\cite{Mauroy_TAC_2016,Susuki_ITPS_2012,Sootla_TAC_2017} for stability analysis, \cite{Korda_Automatica_2018,Sootla_TAC_2018,Sootla_Automatica_2018} for control design, \cite{Surana_CDC_2016,Kaiser_CDC_2018,Proctor_SIAM_2018} for estimation, and \cite{Mauroy_SIAM_2017,Mauroy_TAC_2020} for system identification.}

In this work,  we examine how observability of nonlinear systems can be approached
from a Koopman operator theoretic perspective. Of particular interest to us is how {discrete} symmetries
of the underlying nonlinear system effects the spectral properties of the corresponding 
Koopman operator and how these spectral properties mirror their finite dimensional analogue 
as examined in the context of networked systems~\cite{MEbook}.
However prior to detailing these connections, we examine a canonical representation of the Koopman operator that proves to be useful in our subsequent analysis.
\subsection{Koopman Operator and Infinite-Dimensional Linear Systems} \label{Subsection_Infinite_Dim_Sys}
Koopman eigenfunctions are invariant directions of the dynamics.
As such, we can consider representing the dynamics of the state $\bm{x}$ within the span of the Koopman set $\bm{\Psi}$. In this direction, {recall ${\bm v_j}$'s are Koopman modes associated with observable $\Upsilon({\bf x}) = {\bf x}$ and we propose a new transformation, inspired by the so-called
Koopman Canonical Transform introduced in \cite{Surana_CDC_2016}; the transformation 
assumes the form,
\begin{equation}\label{eq_transformation}
T \left( \bm{x} \right) = 
\begin{bmatrix}
 \phi_1 \circ \bm{x} \\
\phi_2 \circ \bm{x} \\
\cdots
\end{bmatrix},
\end{equation}
where the observation functions are defined using the Koopman eigenfunctions {$\psi_i \in \bm{\Psi}$} as,
\begin{equation*}\label{eq_transformation_modes}
 \begin{cases}
      \phi_i   \left( \bm{x} \right) =  {\bm v_i} \psi_i  \left( \bm{x} \right), & \text{if}\ \lambda_i \in \mathbb{R} \\
      \begin{bmatrix}
\phi_i  \left( \bm{x} \right) \\
\phi_{i+1}  \left( \bm{x} \right) 
\end{bmatrix} =
\begin{bmatrix}
{\mbox{\bf Re}}\, \left( {\bm v_i} \psi_i  \left( \bm{x} \right) \right) \\
{\mbox{\bf Im}} \, \left( {\bm v_i} \psi_i  \left( \bm{x} \right) \right) 
\end{bmatrix}, & \text{if}\ \lambda_i \in \mathbb{C}
    \end{cases};
\end{equation*}
the indexing convention also assumes that $\lambda_{i+1}= \bar{\lambda}_i$ when $\lambda_i \in \mathbb{C}$.}
%
We note that the transformation $T \left( \bm{x} \right)$ is constructed only on the discrete part of the
Koopman spectrum.
%
{The updated state representation $$\bm{z} = T \left( \bm{x} \right),$$ can reconstruct the original state $\bm{x}$ via the Koopman modes,
\begin{equation}\label{eq_reverse_transformation}
\bm{x} (t) =  {\bm V}z(t),
\end{equation}
where the blocks of columns of $\bm{V} $ are,
\begin{equation*}\label{eq_transformation_function}
 \begin{cases}
      \nu_i   = \bm{I}_n, & \text{if}\ \lambda_i \in \mathbb{R} \\
      \begin{bmatrix}
\nu_i  \\
\nu_{i+1}
\end{bmatrix} =
\begin{bmatrix}
\bm{I}_n \\
 \bm{0}
\end{bmatrix}, & \text{if}\ \lambda_i \in \mathbb{C}.
    \end{cases}
\end{equation*}
Note that in our notation every entry of ${\bm z}(t)$ is an element of $\mathbb R^n$. Moreover, 
in light of assumption~\eqref{eq_obs}, $z \in \ell^2(\mathbb R)$.}
This transformation is of interest in our subsequent observability analysis 
as it converts the nonlinear system (\ref{eq_auto_system}) to an infinite-dimensional linear system of the form (\ref{eq_auto_system_Koopman}),
{\begin{equation}\label{eq_auto_system_Koopman}
\begin{split}
 \dot{\bm{z}}(t)& = \left( \bm{\Lambda} \otimes  \bm{I}_n \right) \bm{z}(t),\\
\bm{y} (t) & = h \left( \bm{V} \bm{z}(t) \right),
\end{split}
\end{equation}
with $\bm{z}_0$, the initial condition of $\bm{z}(t)$, belongs to a real, separable infinite-dimensional Hilbert space $\mathcal{Z} \coloneqq \ell_2(\mathbb R)$}, and ${\bm \Lambda}$ is a block diagonal matrix defined according to,
\begin{equation*}\label{eq_transformation_eigenvalues}
 \begin{cases}
      {\bm \Lambda}_{i,i}   = \lambda_i , & \text{if}\ \lambda_i \in \mathbb{R} \\
\begin{aligned}
&
\begin{bmatrix}
{\bm \Lambda}_{i,i} &  {\bm \Lambda}_{i,i+1} \\
{\bm \Lambda}_{i+1,i} &  {\bm \Lambda}_{i+1,i+1}
\end{bmatrix} = \\
&
~~~~~~~~~~~ \vert  \lambda_i \vert
\begin{bmatrix}
 \cos( \angle \lambda_i )  &  \sin( \angle \lambda_i ) \\
- \sin( \angle \lambda_i )  &   \cos( \angle \lambda_i )
\end{bmatrix}
\end{aligned}
& \text{if}\ \lambda_i \in \mathbb{C}
    \end{cases}.
\end{equation*}
{Theorem 2.1.1 in~\cite{singh_book_1993} examines conditions
under which operators such as ${\bm \Lambda}$ are bounded;
furthermore, Proposition 2.3 in~\cite{Fattorini_SIAM_1966} considers reducing an unbounded operator
to a bounded one using its resolvent.
In a nutshell, approximate observability of infinite-dimensional linear systems (\ref{eq_auto_system_Koopman}) and (\ref{eq_auto_system_bilinear}) with an unbounded operator ${\bm \Lambda}$ can be reduced to those with a bounded operator~\cite{Triggiani_SIAM_1975,Triggiani_SIAM_1976}.
As such, in our subsequent analysis boundedness of the operator ${\bm \Lambda}$ will be assumed. 
We also note that in the context of infinite dimensional \textit{linear} systems, the (approximately) observability concepts are dual to those of (approximately) controllability; see Lemma 4.1.13 in \cite{Curtain_Book_2012} and \cite{delfour1972controllability}. 
}
{The system (\ref{eq_auto_system_Koopman}) admits a unique ``mild" solution $\bm{z}(t)$ with the initial condition $\bm{z}_0$ satisfying,
\begin{equation*}\label{eq_koopman_solution}
\bm{z}(t) = \bm{S}(t) \bm{z}_0,
\end{equation*}
where $\bm{S}(t)$ is a continuous semigroup generated by a bounded linear operator $\bm{\Lambda}$ \cite{Curtain_Book_1978,van1996asymptotic}.
}

Since the state measurement is assumed to lie in {the closure of} the span of the Koopman eigenfunctions, we are able to make a connection between the observability of (\ref{eq_auto_system}) and its representation as an infinite-dimensional linear system (\ref{eq_auto_system_Koopman}).
In particular, the measurement function $h$ can be expanded in terms of the (linearly independent)
Koopman eigenfunctions {$\psi_i \in \bm{\Psi}$} as,
{\begin{equation}\label{eq_spans_2}
h \left( \bm{V} \bm{z}(t) \right) = \left( \bm{C} \otimes \mathbbm{1}^\top \right) \bm{z} (t) = 
\begin{bmatrix}
\langle \bm{z} (t) , \bm{c}_1 \otimes \mathbbm{1}^\top \rangle \\
\cdots  \\
\langle \bm{z} (t) , \bm{c}_q \otimes \mathbbm{1}^\top \rangle
\end{bmatrix} ,
\end{equation}
where $\bm{c}_j \otimes \mathbbm{1}^\top$ is the $j$-th row of $\bm{C} \otimes \mathbbm{1}^\top : \mathcal{Z} \rightarrow \mathbb{R}^q$}, for $j=1,\ldots,q$.
We now note that replacing (\ref{eq_spans_2}) for (\ref{eq_auto_system_Koopman}) results in the infinite-dimensional linear system,
{\begin{equation}\label{eq_auto_system_bilinear}
\begin{split}
 \dot{\bm{z}}(t) ~&=~ \left( \bm{\Lambda} \otimes  \bm{I} \right)   \bm{z}(t),\\
 \bm{y} (t)  ~&=~  \left( \bm{C} \otimes \mathbbm{1}^\top \right) \bm{z} (t).
\end{split}
\end{equation}
}
As such, we proceed to analyze the observability of (\ref{eq_auto_system_bilinear}),
and subsequently make a connection between its observability and that of the original system (\ref{eq_auto_system}).
In order to discuss the observability of the infinite-dimensional linear system (\ref{eq_auto_system_bilinear}), we consider the approximate observability as defined in \cite{Curtain_Book_2012}.
\vspace{.05in}
\begin{definition}
The infinite-dimensional linear system (\ref{eq_auto_system_bilinear}) is approximately observable on $[0,\tau ]$, for some finite $\tau > 0$, if knowledge of the measurement function $\bm{y}$ {in $L^2 \left([0,\tau ] ; \bm{y} \right)$\footnote{$L^2 \left([0, \tau]; \bm{y}\right)$ denotes the set of Lebesgue measurable functions with $\int_0^{\tau} \|f(t)\|_2^2 dt < +\infty$.}} uniquely determines its initial state.
\end{definition}
In a nutshell, approximate observability is the ability to estimate initial state based on the knowledge of the measurement data over a finite time interval.
\section{Observability Measures} \label{Section_Observability}

Consider the operator $\bm{\Lambda}$ characterizing the infinite-dimensional linear system (\ref{eq_auto_system_bilinear}) having a discrete spectrum consisting of isolated (countable) Koopman eigenvalues $\lambda_i$, each with multiplicity $r_i$, for $i \in \mathbb{N}$. 
{The operator $\bm {\Lambda}$ admits a complete set of orthonormal eigenvectors $\bm{w}_{ij}$ corresponding to eigenvalues $\lambda_i$} for $i \in \mathbb{N}$ and $j=1,\ldots,r_i$ (using the Gram-Schmidt process) \cite{Halmos_Book_1957}.
If all nonzero eigenvalues have finite multiplicity, the corresponding semigroup $\bm{S}(t)$ can be expanded as {$$\bm{S}(t)\bm{z}(0) =\sum_{i=1}^\infty e^{\lambda_i t} \sum_{j=1}^{r_i} \psi_{ij} \left( \bm{z}_0 \right) \bm{w}_{ij},$$ where $\bm{z}_0$ is the initial condition and $\bm{w}_{ij}$'s are eigenvectors of $\bm{\Lambda}$ and $\psi_{ij} \left( \bm{z}_0 \right) = \langle \bm{z}_0,\bm{w}_{ij} \rangle$ \cite{Klamka_Book_1991,Book_separable_Hilbert_2009}}.
We now note that the functional analytic theory of infinite-dimensional linear systems (on a separable space) leads to a criteria for approximate observability in the Koopman space.
In this direction, we consider a set of {$q \times \left( n r_i \right)$-dimensional constant matrices $\mathcal{O}_i$ 
of the form, for $i \in \mathbb{N}$,
\begin{equation}\label{eq_obser_matrix}
\mathcal{O}_i = 
\begin{bmatrix}
\langle \bm{w}_{i1} , \bm{c}_1 \otimes \mathbbm{1}^\top \rangle & \cdots & \langle \bm{w}_{i r_i} , \bm{c}_1 \otimes \mathbbm{1}^\top \rangle \\
\vdots & \vdots & \vdots \\
\langle \bm{w}_{i1} , \bm{c}_q \otimes \mathbbm{1}^\top \rangle  & \cdots & \langle \bm{w}_{i r_i} , \bm{c}_q \otimes \mathbbm{1}^\top \rangle
\end{bmatrix},
\end{equation}
constructed} in order to investigate the observability problem.
The following result provides a necessary and sufficient condition for approximate observability of an infinite-dimensional linear system (\ref{eq_auto_system_bilinear}) based on the spectral decomposition method discussed in \cite{Klamka_Book_1991,Son_JMAA_1990,Curtain_Book_2012}.
\vspace{0.05in}
\begin{lemma} \label{Lemma_observability}
The system (\ref{eq_auto_system_bilinear}) is approximately observable (over a finite time interval) if and only if for all $i \in \mathbb{N}$,
{\begin{equation}\label{eq_condition_obser}
\text{\bf rank}(\mathcal{O}_i) = n r_i.
\end{equation}
}
\end{lemma}
\begin{proof-non}
{It is known that (\ref{eq_auto_system_bilinear}) is unobservable if there is an eigenvalue of infinite (or arbitrarily high) multiplicity~\cite{fattorini1967complete,Triggiani_SIAM_1976,knowles_Book_1981}.
On the other hand, when the highest eigenvalue multiplicity is finite,~(\ref{eq_condition_obser}) is essentially the generalization of the familiar rank condition for observability of linear finite-dimensional systems. The proof of this generalization has been proposed for linear infinite-dimensional systems in Theorem 5.3 in \cite{Triggiani_SIAM_1976} and Theorem 4.2.1 in \cite{Curtain_Book_2012}. 
}
\end{proof-non}
We now provide a condition for the observability of the nonlinear system (\ref{eq_auto_system}) based on the representation (\ref{eq_auto_system_bilinear}).
\vspace{.05in}
\begin{theorem} \label{Theorem_obs_original_sys}
  {Suppose that the full-state observable ${\bm \Upsilon}\big( {\bm x}\big)$ is in the span of Koopman eigenfunctions.}
Then the nonlinear system (\ref{eq_auto_system}) is observable if and only if for all $i \in \mathbb{N}$,
{\begin{equation}\label{eq_condition_obser_origi}
\text{\bf rank}(\mathcal{O}_i) = n r_i.
\end{equation}
}
\end{theorem}
\begin{proof-non}
Let us assume that there exists $i \in \mathbb{N}$ for which condition (\ref{eq_condition_obser_origi}) is not satisfied.
{Lemma} \ref{Corollary_number_input} thus implies that the system (\ref{eq_auto_system_bilinear}) is not approximately observable.
Since (\ref{eq_auto_system_bilinear}) is diagonal, at least one of the Koopman eigenfunctions associated to $\lambda_i$ can not be uniquely identified by the measurements.
Therefore, the state $\bm{x}$ can not be uniquely constructed via (\ref{eq_obs}), as eigenfunctions are independent and Koopman modes are assumed to be nonzero.
As such, the original system (\ref{eq_auto_system}) is not approximately observable following (\ref{eq_reverse_transformation}).

Now, let us assume that condition (\ref{eq_condition_obser_origi}) is satisfied for all $i \in \mathbb{N}$; that is, based on Lemma \ref{Lemma_observability}, the system (\ref{eq_auto_system_bilinear}) is approximately observable.
According to the diagonal structure of the system (\ref{eq_auto_system_bilinear}), all eigenfunctions are approximately observable.
Therefore, the full-state $\bm{x}(t)$ is approximately observable following the transformation (\ref{eq_reverse_transformation}), as the Koopman modes are nonzero.
Since the approximate observability is the same as observability for finite-dimensional systems, {according to Theorem 17 in  \cite{Boscain_CMP_2015},} the original system (\ref{eq_auto_system}) is observable,
thus completing the proof.
\end{proof-non}

The necessary condition of {Theorem \ref{Theorem_obs_original_sys}} indicates that the number of independent measurements $q$, must be at least equal to the maximum multiplicity of eigenvalues of the operator {$\bm{\Lambda}$}.
Consequently, $\sup_{i=1,2,\ldots} r_i$ plays an important role in investigating the {observability of the nonlinear system (\ref{eq_auto_system})}.
\vspace{.05in}
\begin{corollary} \label{Corollary_number_input}
{The nonlinear system (\ref{eq_auto_system}) is not} observable if
$q < \sup_{i=1,2,\ldots} r_i.$
\end{corollary}
\begin{proof-non}
According to Lemma \ref{Lemma_observability} {and Theorem \ref{Theorem_obs_original_sys}, the nonlinear system (\ref{eq_auto_system})} is unobservable if $\sup_{i=1,2,\ldots} r_i = \infty$.
Let us thus assume that $q < \sup_{i=1,2,\ldots} r_i < \infty$;
hence there exists an eigenvalue $\lambda_i$ for which {$\text{rank}(\mathcal{O}_i) \leq n q < n r_i$.}
The proof now follows by noting that  {condition (\ref{eq_condition_obser_origi}) in Theorem \ref{Theorem_obs_original_sys}} is not satisfied.
\end{proof-non}
Corollary \ref{Corollary_number_input} and Theorem \ref{Theorem_obs_original_sys} can be
used to determine a set of measurements that facilitate ``observing" the system state.
In addition, the structure of the observability matrix (\ref{eq_obser_matrix}) and the operator $\bm{\Lambda}$ provide new insights into how symmetries in the system measurements can lead to unobservability,
a topic we examine next.
\section{{Discrete Symmetries}, Koopman Spectra, and Observability} \label{Section_Symmetry_Koopman}
Analyzing nonlinear systems in an infinite-dimensional setting using independent Koopman eigenfunctions leads to effective means of characterizing connections between symmetry and nonlinear measures of observability.
In this section, we examine structural properties of Koopman eigenvalues, eigenfunctions, and modes of a ``symmetric" nonlinear system (\ref{eq_auto_system}).

We call the {dynamic} system (\ref{eq_auto_system}) {\emph{state symmetric}\footnote{\emph{Note that state symmetry} only concerns the state dynamics $\dot{{\bm x}}(t) = f(x(t))$ without any constraints on the observable ${\bm y}(t)= {\bm h}({\bm x}(t))$.} if there exists a nontrivial permutation matrix $\bm{P}: \mathcal{M} \rightarrow \mathcal{M}$ such that
\begin{equation}\label{symmetry}
f \big( \bm{P} \bm{x} \big) = \bm{P} f \big( \bm{x} \big),
\end{equation}
where $\bm{P}^{k} = \bm{I}$, for some $k \in \{ 2,3,\ldots \}$} {\cite{Aguilar_ACC_2014,Salova_APS_2018,Letellier_Chaos_2002}.} 
{Moreover, we refer to the nonlinear system (\ref{eq_auto_system}) as {\emph{symmetric} if there exists a nontrivial permutation matrix $\bm{P}: \mathcal{M} \rightarrow \mathcal{M}$ such that
\begin{equation}\label{symmetry}
f \big( \bm{P} \bm{x} \big) = \bm{P} f \big( \bm{x} \big),~h \left(\bm {P} \bm{x} \right)= h \left( \bm{x} \right),
\end{equation}
where $\bm{P}^{k} = \bm{I}$, for some $k \in \{ 2,3,\ldots \}$;}
a system is called asymmetric if no such nontrivial permutation exists.\footnote{We note that one can generalize this by allowing general group actions on $\mathcal{M}$ beyond the symmetric group (of permutations). For the purposes of this paper however, we are primary interested in symmetries induced by the automorphism group of a nonlinear network.}
In this section, we examine how {discrete} symmetries in the nonlinear system is reflected in its Koopman operator.
\vspace{.05in}
\begin{theorem} \label{Theorem_Koopman_symmetry}
The {dynamic} system (\ref{eq_auto_system}) is symmetric (with respect to a nontrivial permutation matrix $\bm{P}$) if and only if 
having $\psi_i \big( \bm{x} \big)$ as the Koopman eigenfunction associated with eigenvalue $\lambda_i$ implies that
$\psi_i \big( \bm{P} \bm{x} \big)$ is also a Koopman eigenfunction associated with the same Koopman eigenvalue, for $i \in \mathbb{N}$.
\end{theorem}
\begin{proof-non}
Let us assume that $\psi_i  \big( \bm{x} \big) $ is the Koopman eigenfunction corresponding to the Koopman eigenvalue $\lambda_i$.
Since $$\big \langle \nabla \psi_i  \big( \bm{x} \big),  f \big( \bm{x} \big) \big \rangle  = \lambda_i \psi_i  \big( \bm{x} \big),$$
replacing $\bm{x}$ by $\bm {P} \bm{x}$ results in,
\begin{equation*}
\begin{split}
\big \langle \nabla \psi_i  \big( \bm {P} \bm{x} \big) , f \big( \bm {P} \bm{x} \big) \big \rangle  = \big \langle \nabla \psi_i  \big( \bm {P} \bm{x} \big) , \bm {P} f \big( \bm{x} \big) \big \rangle = \lambda_i \psi_i  \big( \bm {P} \bm{x} \big) .
\end{split}
\end{equation*}
By subtracting the two sides of the above identity from $\big \langle \nabla \psi_i  \big( \bm{x} \big),  f \big( \bm{x} \big) \big \rangle  = \lambda_i \psi_i  \big( \bm{x} \big)$, we obtain,
\begin{equation*} \label{eq_new_eigenfunction}
\begin{split}
\big \langle \nabla \psi_i  \big( \bm{x} \big),  f \big( \bm{x} \big) \big \rangle -& \big \langle \bm {P}^\top \nabla \psi_i  \big( \bm {P} \bm{x} \big) , f \big( \bm{x} \big) \big \rangle
\\
& = \lambda_i \psi_i  \big(\bm{x} \big) - \lambda_i \psi_i  \big( \bm {P} \bm{x} \big) 
 \end{split}
 \end{equation*}
 implying that
 \begin{equation*}
 \begin{split}
\Big \langle \nabla \Big( \psi_i  \big(\bm{x} \big) - \psi_i  \big( \bm {P} \bm{x} \big) \Big) ,  f \big( \bm{x} \big) \Big \rangle   &= \lambda_i \Big( \psi_i  \big(\bm{x} \big) - \psi_i  \big( \bm {P} \bm{x} \big) \Big).
\end{split}
\end{equation*}
Hence  $\psi_i  \big(\bm{x} \big) - \psi_i  \big( \bm {P} \bm{x} \big)$ is a Koopman eigenfunction associated Koopman eigenvalue $\lambda_i$.
Since $\psi_i  \big(\bm{x} \big)$ and $\psi_i  \big(\bm{x} \big) - \psi_i  \big( \bm {P} \bm{x} \big)$ are both Koopman eigenfunction with the same eigenvalue $\lambda_i$, $\psi_i  \big( \bm {P} \bm{x} \big)$ is also a Koopman eigenfunction associated with this Koopman eigenvalue.

Next we show that the nonlinear system is symmetric if $\psi_i \big( \bm{P} \bm{x} \big)$ is also a Koopman eigenfunction associated with the Koopman eigenvalue $\lambda_i$, when $\psi_i ( \bm{x})$ is the Koopman eigenfunction with the same eigenvalue $\lambda_i$, for $i \in \mathbb{N}$.
We first note that $\psi_i  \big(\bm {P} \bm{x} \big)+\psi_i  \big( \bm{x} \big) $, as a linear combination of two eigenfunctions, is itself an eigenfunction of the Koopman operator.
Accordingly, 
\begin{equation*}
\begin{split}
\Big \langle \nabla \Big( \psi_i  \big(\bm {P} \bm{x} \big)+\psi_i  \big( \bm{x} \big) \Big), f \big( \bm{x} \big) \Big \rangle  
&= \lambda_i \Big( \psi_i  \big(\bm {P} \bm{x} \big)+\psi_i  \big( \bm{x} \big) \Big),
\end{split}
\end{equation*}
implying that,
\begin{equation*}
\begin{split}
\big \langle \bm {P}^\top \nabla \psi_i  \big( \bm {P} \bm{x} \big) , f \big( \bm{x} \big) \big \rangle &+ \big \langle \nabla \psi_i  \big( \bm{x} \big),  f \big( \bm{x} \big) \big \rangle 
\\
&= \lambda_i \psi_i  \big( \bm {P} \bm{x} \big)+ \lambda_i \psi_i  \big(\bm{x} \big).
\end{split}
\end{equation*}
By subtracting the above equation from $$\nabla \psi_i  \big( \bm {P} \bm{x} \big)^\top f \big( \bm {P} \bm{x} \big) = \lambda_i \psi_i  \big( \bm {P} \bm{x} \big)$$ and $\big \langle \nabla \psi_i  \big( \bm{x} \big),  f \big( \bm{x} \big) \big \rangle  = \lambda_i \psi_i  \big( \bm{x} \big)$, we conclude that $\big \langle \nabla \psi_i  \big( \bm{x} \big),  \bm{P} f \big( \bm{x} \big) -   f \big(  \bm {P} \bm{x} \big) \big \rangle = \bm{0} $, for $i \in \mathbb{N}$.
{As the full-state observation function $\Upsilon({\bm{x}}) = \bm{x}$ lies in the algebraic span of Koopman eigenfunctions, the \emph{identity matrix}, i.e., Jacobian of $\Upsilon(\bm{x})$, lies in the span of the gradient of the Koopman eigenfunctions. In other words, $\{\nabla \psi_i(\bm{x})\}$ spans $\mathbb R^n$ for every $\bm{x} \in \mathcal M$. Consequently, $\bm{P} f(\bm{x}) - f( \bm{P} \bm{x}) = 0$, i.e., $f \big(  \bm{x} \big) $ is symmetric with respect to $ \bm {P}$, thereby completing the proof.
} 
\end{proof-non}
Theorem \ref{Theorem_Koopman_symmetry} states that the presence of symmetry in the nonlinear system is reflected in the structure of its Koopman eigenfunctions.
In this case, the ``projected symmetry" of a Koopman eigenfunction is either along the same direction as the original eigenfunction or along a new distinct direction.
The notions of rotational and ref lectional symmetry further {clarify} this distinction.

The Koopman eigenvalue/eigenfunction pair $(\lambda_i, \psi_i)$ is said to have a {\em rotational} symmetry if the action of the symmetry (from the dynamics) on the Koopman eigenfunction leads to a linearly dependent Koopman eigenfunction, i.e., 
\begin{equation}\label{eq_Cond_Eigenfunction_1}
\psi_i  \big( \bm{x} \big) = c \psi_i  \big( \bm{P} \bm{x} \big),
\end{equation}
where $c \in \mathbb{C}$.
On the other hand, we refer to the Koopman eigenfunction
as {\em reflectional} when
 the action of the symmetry on this eigenfunction leads to a linearly independent Koopman eigenfunction (with respect to the original one).
 In this case, there exists another Koopman eigenfunction $\psi_j ( \bm{x})$, not along $\psi_i (\bm{x})$, 
 with the same Koopman eigenvalue $\lambda_i$, such that
\begin{equation}\label{eq_Cond_Eigenfunction_2}
\psi_j  \big( \bm{x} \big) = c \psi_i  \big( \bm{P} \bm{x} \big),
\end{equation}
where $c \in \mathbb{C}$.
Hence, Theorem \ref{Theorem_Koopman_symmetry} states that the presence of symmetry in the nonlinear system leads to either rotational and reflectional symmetry in the Koopman eigenfunctions.
\vspace{.05in}

\begin{lemma} \label{Lemma_pair_reflectional}
Suppose that the dynamic system (\ref{eq_auto_system}) is symmetric and {the full-state observation vector and the measurement are in the span of the Koopman set}. Then if the Koopman set includes a reflectional eigenfunction, the ``reflected" eigenfunction
belongs to the Koopman set.
\end{lemma}
\begin{proof-non}
Without loss of generality, let us assume that the eigenfunctions $\psi_{f1} \big(\bm{x} \big)$, $\psi_{f2} \big(\bm{x} \big)$, $\ldots$ $\in \bm{\Psi},$ with the associated eigenvalues $\lambda_{f1} $, $ \lambda_{f2} $, $ \ldots,$ have rotational symmetry, that is, $\psi_{fi} \big( \bm{P} \bm{x} \big) = {c_{fi}} \psi_{fi} \big(\bm{x} \big) $, and $\psi_{r1} \big(\bm{x} \big)$, $ \psi_{t1} \big(\bm{x} \big)$, $\psi_{r2} \big(\bm{x} \big)$, $ \psi_{t2} \big(\bm{x} \big)$, $\ldots$ $\in \bm{\Psi},$ with the associated eigenvalues $\lambda_{r1}$, $ \lambda_{t1}$, $\lambda_{r2}$, $ \lambda_{t2}$, $ \ldots,$ have the reflectional symmetry such that $\psi_{ri} \big(\bm{P} \bm{x} \big) = {c_{ri}}  \psi_{ti} \big( \bm{x} \big)$, $\psi_{ti} \big(\bm{P} \bm{x} \big) =  {c_{ti}}  \psi_{ri} \big(\bm{x} \big)$, $\lambda_{ri} = \lambda_{ti}$, and reflectional eigenfunctions $\psi_{n1} \big(\bm{x} \big) \in \bm{\Psi}$ with the associated eigenvalues $\lambda_{n1}$, while $\psi_{n1} \big(\bm{P} \bm{x} \big) \notin \bm{\Psi}$, {and $c_{fi},c_{ti},c_{ri}  \in \mathbb{C}$}.
Then, the state $\bm{x}$ can be represented as,
{\begin{equation*}
\begin{split}
\bm{x}(t) =& 
e^{\lambda_{n1}} \psi_{n1} \big( \bm{x} \big) \bm{v}_{nj}
+\sum_{j=1}^\infty e^{\lambda_{fj}} \psi_{fj} \big( \bm{x} \big) \bm{v}_{fj} 
 \\
&
+ \sum_{i=1}^\infty e^{\lambda_{ti}} \psi_{ti} \big( \bm{x} \big) \bm{v}_{ti} 
+ \sum_{i=1}^\infty e^{\lambda_{ri}} \psi_{ri} \big( \bm{x} \big) \bm{v}_{ri}.
\end{split}
\end{equation*}}
Replacing $\bm{x}$ by $\bm {P} \bm{x}$ and using rotational and reflectional symmetry of the eigenfunctions result in,
\begin{equation*}
\begin{split}
\bm{P} \bm{x}(t) =& 
{e^{\lambda_{n1}}} \psi_{n1} \big( \bm {P} \bm{x} \big) \bm{v}_{n1} 
+ \sum_{j=1}^\infty {e^{\lambda_{fj}}} {c_{fi}} \psi_{fj} \big( \bm{x} \big) \bm{v}_{fj} 
\\
&
+ \sum_{i=1}^\infty {e^{\lambda_{ri}}} {c_{ti}} \psi_{ri} \big( \bm{x} \big) \bm{v}_{ti} 
+ \sum_{i=1}^\infty {e^{\lambda_{ti}}} {c_{ri}} \psi_{ti} \big( \bm{x} \big) \bm{v}_{ri}.
\end{split}
\end{equation*}
In the meantime, left multiplication by $\bm{P}^{k-1}$ leads to another expansion of the 
state $\bm{x}$ as,
\begin{equation*}
\begin{split}
&\bm{x}(t) ={e^{\lambda_{n1}}} \psi_{n1} \big( \bm {P} \bm{x} \big) \bm{P}^{k-1} \bm{v}_{n1} 
\\
&
+ \sum_{j=1}^\infty {e^{\lambda_{fj}}} \psi_{fj} \big( \bm{x} \big) {c_{fi}}  \bm{P}^{k-1} \bm{v}_{fj} 
+ \sum_{i=1}^\infty {e^{\lambda_{ri}}} \psi_{ri} \big( \bm{x} \big) { c_{ti}} \bm{P}^{k-1} \bm{v}_{ti}
\\
&
+ \sum_{i=1}^\infty {e^{\lambda_{ti}}} \psi_{ti} \big( \bm{x} \big) { c_{ri}} \bm{P}^{k-1} \bm{v}_{ri}.
\end{split}
\end{equation*}
Consequently, the full state $\bm{x}$ is in the span of the set $\psi_{n1} \big( \bm {P} \bm{x} \big) \cup \bm{\Psi} \setminus \psi_{n1} \big( \bm{x} \big) $.
Therefore, $\psi_{n1} \big( \bm {P} \bm{x} \big)$ is in the span of $\bm{\Psi}$ and 
\begin{equation} \label{eq_phi_px}
\begin{split}
\psi_{n1} \big( \bm {P} \bm{x} \big)  =& 
\mu_{n1} \psi_{n1} \big( \bm{x} \big) 
+\sum_{j=1}^\infty \mu_{fj} \psi_{fj} \big( \bm{x} \big)
 \\
&
+ \sum_{i=1}^\infty \mu_{ti} \psi_{ti} \big( \bm{x} \big) 
+ \sum_{i=1}^\infty \mu_{ri} \psi_{ri} \big( \bm{x} \big),
\end{split}
\end{equation}
where there exist $\mu_{fj} \neq 0$, $\mu_{ti} \neq 0$, or $\mu_{ri} \neq 0$ since $\psi_{n1} \big( \bm{x} \big)$ and $\psi_{n1} \big( \bm {P} \bm{x} \big)$ are linearly independent.
Replacing $\bm{x}$ by $\bm {P} \bm{x}$ in equation (\ref{eq_phi_px}) and applying equation (\ref{eq_phi_px}) result in
\begin{equation*}
\begin{split}
& \psi_{n1} \big( \bm {P}^2 \bm{x} \big)  =
\mu_{n1}^2 \psi_{n1} \big( \bm{x} \big) 
+\sum_{j=1}^\infty \mu_{fj} (\mu_{n1}+c_{fj}) \psi_{fj} \big( \bm{x} \big)
 \\
&
+ \sum_{i=1}^\infty \mu_{ti} (\mu_{n1}+c_{ri})  \psi_{ti} \big( \bm{x} \big) 
+ \sum_{i=1}^\infty \mu_{ri} (\mu_{n1}+c_{ti})  \psi_{ri} \big( \bm{x} \big).
\end{split}
\end{equation*}

By repeating the above procedure $k-1$ times, replacing $\bm{x}$ by $\bm {P} \bm{x}$ and applying (\ref{eq_phi_px}) and considering $\bm {P}^k= \bm {I}$, result in
\begin{equation*}
\begin{split}
\psi_{n1} \big( \bm {P}^k \bm{x} \big)  =&  \psi_{n1} \big( \bm{x} \big)  =
\xi_{n1} \psi_{n1} \big( \bm{x} \big) 
+\sum_{j=1}^\infty \xi_{fj} \psi_{fj} \big( \bm{x} \big)
 \\
&
+ \sum_{i=1}^\infty \xi_{ti} \psi_{ti} \big( \bm{x} \big) 
+ \sum_{i=1}^\infty \xi_{ri} \psi_{ri} \big( \bm{x} \big),
\end{split}
\end{equation*}
where $\xi_{n1}$, $\xi_{fj}$, $\xi_{ti}$, and $\xi_{ri}$ are functions of $\mu_{n1}$, $\mu_{fj}$, $\mu_{ti}$, $\mu_{ri}$, $c_{fj}$, $c_{ti}$, and $c_{ri}$.
Therefore, $\psi_{n1} \big( \bm{x} \big)$ lies in the span of the set $\bm{\Psi} \setminus \psi_{n1} \big( \bm{x} \big)$.
This is in contradiction with the independence assumption of the Koopman set $\bm{\Psi}$.
\end{proof-non}
{We now proceed to show how symmetries in a dynamic system are reflected in the corresponding Koopman modes.}

\vspace{.05in}
\begin{corollary} \label{Corollary_Koopman_symmetry}
Suppose that the dynamic system (\ref{eq_auto_system}) is symmetric and {the full-state observation vector and the measurement are in the span of the Koopman set}. Then, the following statements hold for some $c_{fi},c_{ri} \in \mathbb{C}$,
\begin{enumerate}[label=\alph*)]
\item
if there exists a rotational symmetric {eigenfunctions} $\psi_i$ in the Koopman set, then $\bm{v}_i = c_{fi} \bm{P}^{k-1} \bm{v}_i$, where  $\bm{v}_i$ is its associated Koopman mode.
\item 
if there exists a pair of reflectional symmetric {eigenfunctions} $\psi_i$ and $\psi_j $ in  the Koopman set, then $\bm{v}_j  = c_{ri} \bm{P}^{k-1} \bm{v}_i$, where $\bm{v}_i, \bm{v}_j$ are the associated Koopman modes.
\end{enumerate}

\end{corollary}
\begin{proof-non}
According to Lemma \ref{Lemma_pair_reflectional}, the Koopman set cannot contain only one pair of reflectional symmetry eigenfunctions.
Without loss of generality, let us assume that the eigenfunctions $\psi_{f1} \big(\bm{x} \big)$, $\psi_{f2} \big(\bm{x} \big)$, $\ldots$ $\in \bm{\Psi},$ with the associated eigenvalues $\lambda_{f1} $, $ \lambda_{f2} $, $ \ldots,$ have the rotational symmetry, that is, $\psi_{fi} \big( \bm{P} \bm{x} \big) = {c_{fi}} \psi_{fi} \big(\bm{x} \big) $, and $\psi_{r1} \big(\bm{x} \big)$, $ \psi_{t1} \big(\bm{x} \big)$, $\psi_{r2} \big(\bm{x} \big)$, $ \psi_{t2} \big(\bm{x} \big)$, $ \ldots$ $\in \bm{\Psi},$ with the associated eigenvalues $\lambda_{r1}$, $ \lambda_{t1}$, $\lambda_{r2}$, $ \lambda_{t2}$, $ \ldots,$ have the reflectional symmetry such that $\psi_{ri} \big(\bm{P} \bm{x} \big) = {c_{ri}}  \psi_{ti} \big( \bm{x} \big)$, $\psi_{ti} \big(\bm{P} \bm{x} \big) =  {c_{ti}}  \psi_{ri} \big(\bm{x} \big)$, $\lambda_{ri} = \lambda_{ti}$, for $i \in \mathbb{N}$, {and $c_{fi},c_{ti},c_{ri}  \in \mathbb{C}$}.
Then, the state $\bm{x}$ can be expanded as,
\begin{equation*}
\begin{split}
\bm{x}(t) =& 
\sum_{j=1}^\infty {e^{\lambda_{fj}}} \psi_{fj} \big( \bm{x} \big) \bm{v}_{fj} 
+ \sum_{i=1}^\infty {e^{\lambda_{ti}}} \psi_{ti} \big( \bm{x} \big) \bm{v}_{ti} 
 \\
&
+ \sum_{i=1}^\infty {e^{\lambda_{ri}}} \psi_{ri} \big( \bm{x} \big) \bm{v}_{ri}.
\end{split}
\end{equation*}

Replacing $\bm{x}$ by $\bm {P} \bm{x}$ and using rotational and reflectional symmetry of the eigenfunctions result in,
\begin{equation*}
\begin{split}
\bm{P} \bm{x}(t) =& 
\sum_{j=1}^\infty {e^{\lambda_{fj}}} {c_{fj}} \psi_{fj} \big( \bm{x} \big) \bm{v}_{fj} 
+ \sum_{i=1}^\infty {e^{\lambda_{ri}}} {c_{ti}} \psi_{ri} \big( \bm{x} \big) \bm{v}_{ti} 
\\
&
+ \sum_{i=1}^\infty {e^{\lambda_{ti}}} {c_{ri}} \psi_{ti} \big( \bm{x} \big) \bm{v}_{ri}.
\end{split}
\end{equation*}
In the meantime, left multiplication by $\bm{P}^{k-1}$ leads to another expansion of $\bm{x}$ as,
\begin{equation*}
\begin{split}
&\bm{x}(t) = 
 \sum_{j=1}^\infty {e^{\lambda_{fj}}} \psi_{fj} \big( \bm{x} \big) {c_{fj}}  \bm{P}^{k-1} \bm{v}_{fj} 
\\
&
+ \sum_{i=1}^\infty {e^{\lambda_{ri}}} \psi_{ri} \big( \bm{x} \big) { c_{ti}} \bm{P}^{k-1} \bm{v}_{ti}
+ \sum_{i=1}^\infty {e^{\lambda_{ti}}} \psi_{ti} \big( \bm{x} \big) { c_{ri}} \bm{P}^{k-1} \bm{v}_{ri}.
\end{split}
\end{equation*}

By defining a new set of Koopman modes as $\bm{\bar{v}}_{fj} = c_{fj} \bm{P}^{k-1} \bm{v}_{fj} $, $\bm{\bar{v}}_{ri} = c_{ti} \bm{P}^{k-1} \bm{v}_{ti}$, and $\bm{\bar{v}}_{ti} = c_{ti} \bm{P}^{k-1} \bm{v}_{ri}$, we can now express the state $\bm{x}$ as,
\begin{equation*}
\begin{split}
\bm{x}(t) =& \sum_{j=1}^\infty {e^{\lambda_{fj}}} \psi_{fj} \big( \bm{x} \big) \bm{\bar{v}}_{fj}
+ \sum_{i=1}^\infty {e^{\lambda_{ri}}} \psi_{ri} \big( \bm{x} \big) \bm{\bar{v}}_{ri}
\\
&
+ \sum_{i=1}^\infty {e^{\lambda_{tk}}} \psi_{ti} \big( \bm{x} \big) \bm{\bar{v}}_{ti}.
\end{split}
\end{equation*}
Note that the coefficients associated with the same eigenfunction in the first and last expansions of $\bm{x}$ are identical. Therefore, $\bm{v}_{fj} = c \bm{P}^{k-1} \bm{v}_{fj} $, $\bm{v}_{ti} = c \bm{P}^{k-1} \bm{v}_{ri}$, and $\bm{v}_{ti} = c \bm{P}^{k-1} \bm{v}_{ri}$, thus completing  the proof.
\end{proof-non}

Theorem \ref{Theorem_Koopman_symmetry} and Corollary \ref{Corollary_Koopman_symmetry} highlight how symmetry in a {dynamic} system is reflected in the spectral properties of the 
corresponding Koopman operator.
{The following result 
shows that when the nonlinear system is symmetric, the corresponding infinite-dimensional linear system has repeated eigenvalues.
\vspace{.05in}
\begin{lemma} \label{Lemma_sym_sys}
Suppose that the nonlinear system (\ref{eq_auto_system}) is symmetric (with respect to a nontrivial permutation matrix) and {the full-state observation vector and the measurement are in the span of the Koopman set}. Then, the infinite-dimensional linear {systems (\ref{eq_auto_system_Koopman}) and (\ref{eq_auto_system_bilinear}) have} repeated eigenvalues.
\end{lemma}
\begin{proof-non}
If the Koopman set includes reflectional eigenfunctions, then the proof is completed since reflectional eigenfunctions admit a repeated set of Koopman eigenvalues.
Without loss of generality, let us assume that the Koopman set $\bm{\Psi}$ contains only rotational eigenfunctions $\psi_{f1} \big(\bm{x} \big)$, $\psi_{f2} \big(\bm{x} \big)$, $\ldots,$ with the associated eigenvalues $\lambda_{f1} $, $ \lambda_{f2} $, $ \ldots$, such that $\psi_{fi} \big( \bm{P}  \bm{x} \big) = c_{fi} \psi_{fi} \big(\bm{x} \big)$ and $\lambda_{fi} \neq \lambda_{fj} $, for $i \neq j$ and $c_{fi} \in \mathbb{C}$.
Since the measurement $h \left( \bm{x}(t) \right)$ is in the span of the Koopman set $\bm{\Psi}$, it can be expanded as,
$ h \left( \bm{x} \right) =  
\sum_{j=1}^\infty {e^{\lambda_{fj}}} \psi_{fj} \big( \bm{x} \big) \bm{q}_{fj},
$
where $\bm{q} \in \mathbb{R}^q$. Replacing $\bm{x}$ by $\bm {P} \bm{x}$ and using the symmetry of the measurement ($h \left( \bm{P} \bm{x} \right) =h \left( \bm{x} \right)$) result in,
\begin{equation*}
\begin{split}
h \left( \bm{x} \right) =  h \left( \bm{P} \bm{x} \right)
& =
\sum_{j=1}^\infty {e^{\lambda_{fj}}} \psi_{fj} \big( \bm{P} \bm{x} \big) \bm{q}_{fj}
 \\
&=
\sum_{j=1}^\infty {e^{\lambda_{fj}}} c_{fi} \psi_{fj} \big( \bm{x} \big) \bm{q}_{fj}.
\end{split}
\end{equation*}
Since these Koopman eigenfunctions are linearly independent, we conclude that $c_{fi} = 1$ by comparing the obtained expansion and the original expansion of the measurement equation. 
Therefore, $\psi_{fi} \big( \bm{P}  \bm{x} \big) = \psi_{fi} \big(\bm{x} \big)$.
Now, we expand the state $\bm{x}$ as,
$\bm{x}(t) =
\sum_{j=1}^\infty {e^{\lambda_{fj}}} \psi_{fj} \big( \bm{x} \big) \bm{v}_{fj}.
$ Replacing $\bm{x}$ by $\bm {P} \bm{x}$ and using $\psi_{fi} \big( \bm{P}  \bm{x} \big) = \psi_{fi} \big(\bm{x} \big)$ result in,
\begin{equation*}
\begin{split}
\bm{P} \bm{x}(t)
&= \sum_{j=1}^\infty {e^{\lambda_{fj}}} \psi_{fj} \big(\bm {P} \bm{x} \big) \bm{v}_{fj} 
=\sum_{j=1}^\infty {e^{\lambda_{fj}}} \psi_{fj} \big( \bm{x} \big) \bm{v}_{fj} 
\\
& = \bm{x}(t).
\end{split}
\end{equation*}
Since this identity holds for all $\bm{x}(t)$, we conclude that $\bm{P} = \bm{I}$.
This however is a contradiction, as  $\bm{P}$ is assumed to be a nontrivial permutation.
\end{proof-non}
%
We now observe a commutativity property for nonlinear symmetric systems that is rather analogous to their
linear counterparts. 
\vspace{.05in}
\begin{lemma} \label{Lemma_symmetry_dynamic}
When the nonlinear system (\ref{eq_auto_system}) is symmetric and {the full-state observation vector and the measurement are in the span of the Koopman set}, then the system (\ref{eq_auto_system_bilinear}) is symmetric with respect to a nonidentity operator {$ \bm{Q} \otimes \bm{I}: \mathcal{Z} \rightarrow \mathcal{Z}$, for which $ \left( \bm{Q} \otimes \bm{I} \right) \left( \bm{\Lambda} \otimes \bm{I} \right) = \left( \bm{\Lambda} \otimes \bm{I} \right) \left( \bm{Q} \otimes \bm{I} \right)$ and $ \bm{C} \otimes  \mathbbm{1}^\top= \left(\bm{C} \otimes \mathbbm{1}^\top \right) \left( \bm{Q} \otimes \bm{I} \right)$.}
\end{lemma}
\begin{proof-non}
According to Theorem \ref{Theorem_Koopman_symmetry}, the Koopman eigenfunctions associated with same eigenvalues, are symmetric with respect to some non-trivial permutation $\bm{P}$.
{It thus follows from Lemma \ref{Lemma_sym_sys} that 
the corresponding Koopman set includes reflectional eigenfunctions and the corresponding operator $\bm{\Lambda}$ has repeated eigenvalues.}
Without loss of generality, let the eigenfunctions $\psi_{r1} \big(\bm{x} \big)$, $ \psi_{t1} \big(\bm{x} \big)$, $\psi_{r2} \big(\bm{x} \big)$, $ \psi_{t2} \big(\bm{x} \big)$, $ \ldots$ $ \in \bm{\Psi},$ with the associated eigenvalues $\lambda_{r1}$, $ \lambda_{t1}$, $\lambda_{r2}$, $ \lambda_{t2}$, $ \ldots$ have the reflectional symmetry such that $\psi_{ri} \big(\bm{x} \big) = c_{ri}  \psi_{ti} \big(\bm{P} \bm{x} \big)$, $\psi_{ti} \big(\bm{x} \big) =c_{ti} \psi_{ri} \big(\bm{P} \bm{x} \big)$, $\lambda_{ri} = \lambda_{ti}$, and eigenfunctions $\psi_{f1} \big(\bm{x} \big)$, $\psi_{f2} \big(\bm{x} \big)$, $\ldots$ $ \in \bm{\Psi},$ with the associated eigenvalues $\lambda_{f1} $, $ \lambda_{f2} $, $ \ldots,$ have the rotational symmetry, $\psi_{fi} \big( \bm{P} \bm{x} \big) =c_{fi} \psi_{fi} \big(\bm{x} \big) $.

We now construct the diagonal operator $\bm{\Lambda}$ with diagonal elements $\lambda_{f1} $, $ \lambda_{f2} $, $ \ldots$, $\lambda_{r1}$, $ \lambda_{t1}$, $\lambda_{r2}$, $ \lambda_{t2}$, $ \ldots$.
Define the permutation operator $\bm{Q}$ that exchanges 
the $ri$-th and $ti$-th elements, for $i \in \mathbb{N}$.
Since $\lambda_{ri} = \lambda_{ti}$, $\bm{Q} \bm{\Lambda} = \bm{\Lambda} \bm{Q}$, {and consequently $\left( \bm{Q} \bm{\Lambda} \right) \otimes \bm{I}= \left( \bm{\Lambda} \bm{Q} \right) \otimes \bm{I} \Rightarrow  \left( \bm{Q} \otimes \bm{I} \right) \left( \bm{\Lambda} \otimes \bm{I} \right) = \left( \bm{\Lambda} \otimes \bm{I} \right) \left( \bm{Q} \otimes \bm{I} \right)$.
Thus, the operators $\bm{\Lambda}$ and $\bm{\Lambda}\otimes \bm{I} $ are symmetric with respect to the non-identity permutation $\bm{Q}$ and $\bm{Q}\otimes \bm{I} $, respectively.}

Applying the transformation (\ref{eq_reverse_transformation}) in $h \left(\bm {P} \bm{x} \right)= h \left( \bm{x} \right)$ and taking into account that {$\left( \bm{Q} \otimes \bm{I} \right) \left( \bm{\Lambda} \otimes \bm{I} \right) = \left( \bm{\Lambda} \otimes \bm{I} \right) \left( \bm{Q} \otimes \bm{I} \right)$ results in $h \left(\bm {V} \left( \bm{Q} \otimes \bm{I} \right) \bm{z} \right)= h \left(\bm{V} \bm{z} \right)$.}
This latter identity can now be written based on the expansion (\ref{eq_spans_2}) as {$\left( \bm{C} \otimes  \mathbbm{1}^\top \right) \bm{z} =\left( \bm{C} \otimes  \mathbbm{1}^\top \right) \left( \bm{Q} \otimes \bm{I} \right) \bm{z} = \left( \bm{C} \bm{Q} \otimes \mathbbm{1}^\top \right) \bm{z}  $}.
Since the elements of $\bm{z}(t)$ are nonzero for all times and the operator $\bm{Q}$ is not identity, it follows that $\bm{C} =\bm{C} \bm{Q}$ {and $ \bm{C} \otimes  \mathbbm{1}^\top= \left(\bm{C} \otimes \mathbbm{1}^\top \right) \left( \bm{Q} \otimes \bm{I} \right)$}.
Therefore, the system (\ref{eq_auto_system_bilinear}) is symmetric with respect to {$\bm{Q} \otimes \bm{I}$}.
\end{proof-non}
%
We are now in the position to clarify how {discrete} symmetries in a nonlinear system lead to its unobservability.
\subsection{Role of {Discrete} Symmetries on Observability} \label{Subsection_symmetry_obser}
We now analyze the observability of a {discrete} symmetric nonlinear system.
One of the unique features of our approach is 
utilizing the symmetry in the Koopman representation
of the nonlinear system for such an analysis.
%
This is done by showing that the symmetry in the nonlinear system {induces a multiplicity in the Koopman spectra, leading to unobservability of the system.}
\vspace{.05in}
\begin{theorem} \label{Theorem_Obs_Sys_Syemmetry}
{ Suppose that the nonlinear system (\ref{eq_auto_system}) is symmetric and the full-state observation vector and the measurement are in the span of the Koopman set}; then the system (\ref{eq_auto_system}) is unobservable.
\end{theorem}
\begin{proof-non}
Since the nonlinear system (\ref{eq_auto_system}) is symmetric, Lemma \ref{Lemma_symmetry_dynamic} implies that (\ref{eq_auto_system_bilinear}) is symmetric with respect to matrix {$\bm{Q} \otimes \bm{I}$} and there exists a repeated eigenvalue $\lambda_i$ {for the corresponding $\bm{\Lambda}$} with multiplicity $r_i \geq 2$ and {$ \bm{C} \otimes  \mathbbm{1}^\top= \bm{C} \bm{Q} \otimes \mathbbm{1}^\top = \left(\bm{C} \otimes \mathbbm{1}^\top \right) \left( \bm{Q} \otimes \bm{I} \right)$}. As such, there exists a set of eigenvectors $\bm{w}_{ij}$ associated with the repeated eigenvalue $\lambda_i$ such that {$\left( \bm{Q} \otimes \bm{I} \right) \bm{w}_{ij}$} is also an eigenvector.
Hence, {$\left( \bm{Q} \otimes \bm{I} \right) \bm{w}_{ij} -\bm{w}_{ij}$} is also an eigenvector of the matrix {$\bm{\Lambda} \otimes \bm{I}$} corresponding to the eigenvalue $\lambda_i$, for $j=1,\ldots,r_i$.
However, the eigenvector {$\left( \bm{Q} \otimes \bm{I} \right) \bm{w}_{ij} -\bm{w}_{ij}$} is orthogonal to {$ \bm{C} \otimes  \mathbbm{1}^\top$} as
{\begin{equation*}
\begin{split}
&\langle \left( \bm{Q} \otimes \bm{I} \right)  \bm{w}_{ij} -\bm{w}_{ij} , \bm{C} \otimes  \mathbbm{1}^\top \rangle   \\ 
& \qquad \qquad = \langle \left( \bm{Q} \otimes \bm{I} \right)  \bm{w}_{ij}  , \bm{C} \otimes  \mathbbm{1}^\top \rangle - \langle \bm{w}_{ij} , \bm{C} \otimes  \mathbbm{1}^\top \rangle \\
 &\qquad \qquad = \langle  \bm{w}_{ij}  , \left( \bm{C} \otimes  \mathbbm{1}^\top \right) \left( \bm{Q} \otimes \bm{I} \right)  \rangle - \langle \bm{w}_{ij} , \bm{C} \otimes  \mathbbm{1}^\top \rangle \\
 & \qquad \qquad =  \langle  \bm{w}_{ij}  , \bm{C} \otimes  \mathbbm{1}^\top \rangle - \langle \bm{w}_{ij} , \bm{C} \otimes  \mathbbm{1}^\top \rangle = \bm{0}.
\end{split}
\end{equation*}

This, on the other hand, implies that $\text{\bf rank}(\mathcal{O}_i) < n r_i$.}
Consequently, the system (\ref{eq_auto_system}) is not observable following Theorem \ref{Theorem_obs_original_sys}.
\end{proof-non}
Theorem \ref{Theorem_Obs_Sys_Syemmetry} states that the presence of symmetry in 
the nonlinear system is sufficient for unobservability.
%
%
Our next result pertains to the relation between unobservability and the number of measurements for nonlinear systems.

\vspace{.05in}
\begin{corollary} \label{Corollary_Obs_Auto_Syemmetry}
Suppose that the dynamic system (\ref{eq_auto_system}) is symmetric, {the full-state observation vector and the measurement are in the span of the Koopman set, and the Koopman set} includes {eigenfunctions} with same eigenvalues.
If the maximum multiplicity of a Koopman eigenvalue is greater than the number of measurements,
then (\ref{eq_auto_system}) is unobservable.
\end{corollary}
\begin{proof-non}
Let us assume that the nonlinear {dynamic} system is symmetric and $\lambda_i$ is a repeated eigenvalue of the corresponding $\bm{\Lambda}$ with multiplicity greater than the number of measurements, $r_i > q$.
Since {$\bm{\Lambda} \otimes \bm{I}$} is diagonal (and the underlying Hilbert space is separable), the standard orthonormal basis can be considered as the set of its eigenvectors.
Let us define the matrix {$\bm{E}_i= [ \bm{e}_{i1},\ldots,\bm{e}_{i \left(n r_i\right)} ]$} such that vector $\bm{e}_{ij}$ is an eigenvector of {$\bm{\Lambda} \otimes \bm{I}$ associated with eigenvalue $\lambda_i$, where $\bm{e}_{ij}$ is the unit vector, for $j=1,\ldots,n r_i$.}
Since $r_i > q$, there exist {$n \left( r_i - q \right)$ orthogonal unit vectors $\big\{ \bm{v}_{i1},\ldots,\bm{v}_{i \left(n (r_i-q) \right)} \big\} \in \mathbb{R}^{r_i}$ such that $ \langle \bm{v}_{ij} , [ \langle \bm{e}_{ij} , \bm{c}_k \otimes  \mathbbm{1}^\top \rangle ,\ldots, \langle \bm{e}_{i r_i}, \bm{c}_k \otimes  \mathbbm{1}^\top \rangle ]^\top \rangle = 0$, for $j=1,\ldots,n (r_i-q)$ and $k=1,\ldots,q$.
Orthogonal unit vectors $\big\{ \bm{v}_{i1},\ldots,\bm{v}_{i (n r_i)} \big\}$ are constructed as a basis for $\mathbb{R}_{n r_i}$, where $\big\{ \bm{v}_{i1},\ldots,\bm{v}_{i \left(n (r_i-q)\right)} \big\}$} are now constructed by applying the Gram-Schmidt orthogonalization procedure \cite{Halmos_Book_1957}.
The new set of eigenvectors associated with eigenvalue $\lambda_i$ are thereby obtained as {$$\begin{bmatrix} \bm{w}_{i1},\ldots,\bm{w}_{i \left(n r_i\right)} \end{bmatrix} = \bm{E}_i \begin{bmatrix} \bm{v}_{i1},\ldots,\bm{v}_{i \left( n r_i \right)} \end{bmatrix}.$$
Consequently, $\bm{w}_{ij}$ is orthogonal to $\bm{C} \otimes \mathbbm{1}^\top $, for $j=1,\ldots,\left(n (r_i-q) \right)$ and $\text{\bf rank} (\mathcal{O}_i) = n (r_i-q) < n r_i$}.
The application of Theorem \ref{Theorem_obs_original_sys} now completes the proof.
\end{proof-non}
Corollary \ref{Corollary_Obs_Auto_Syemmetry} states that the minimum number of the measurements needed to make the system observable is the maximum multiplicity of the Koopman eigenvalues.

It is instructive to note that a more streamlined algebraic approach to nonlinear observability can be used to prove Theorem \ref{Theorem_Obs_Sys_Syemmetry} when the underlying symmetry is an involution ($\bm{P}^{2} = \bm{I}$).

{\begin{remark}
The results proposed in Section~\ref{Section_Symmetry_Koopman} are obtained under the assumption that the measurement lie in the span of the Koopman set. This assumption can be satisfied by proper choice of the Koopman set.
\end{remark}}

We note that without the stated assumptions,  Theorem \ref{Theorem_Obs_Sys_Syemmetry} and Corollary \ref{Corollary_Obs_Auto_Syemmetry} are less straightforward to prove.
In the meantime, Corollary \ref{Corollary_Obs_Auto_Syemmetry} has been numerically 
demonstrated through simulation studies in \cite{Whalen_PhysRevX_2015} (where the system with reflectional symmetries and a single measurement is shown to be unobservable).
{Furthermore, Theorem \ref{Theorem_Obs_Sys_Syemmetry} and Corollary \ref{Corollary_Obs_Auto_Syemmetry} clarify the role of repeated Koopman eigenvalues in the observability analysis; in particular, why symmetric nonlinear systems containing only
rotational symmetries may remain observable, while symmetric nonlinear systems with the reflectional symmetry are always unobservable.}
\section{Illustrative examples} \label{Section_Examples}
In this section, we consider three examples that demonstrate the application of the results discussed in the paper. The first example pertains to linear networks over undirected and directed graphs; the second example, pertains to a suitably constructed nonlinear system, and in the third example, we examine the application of the developed theory to a network of nanoelectromechanical systems.\\~\\
{
\textit{Example 1}.
We consider the consensus problem in undirected and directed networks of 3 dynamic agents with topologies shown in Figure \ref{fig_3agents}.
\begin{figure}[!ht]
\begin{center}
\begin{tikzpicture}[->,>=stealth',shorten >=1pt,auto,node distance=1.8cm,
                    semithick]
  \node[state,fill=green!50, text=black, circle,scale=0.6, draw=black] (N1)                               {$1$};
  \node[state,fill=blue!50, text=black, circle,scale=0.6, draw=black] (N2) [below left of=N1]  {$2$};
  \node[state,fill=red!50, text=black, circle,scale=0.6, draw=black] (N3) [below right of=N1] {$3$};
  \node[] (Na) [right of=N3]  {}; 
  \node[state,fill=blue!50, text=black, circle,scale=0.6, draw=black] (Nd2) [right of=Na]  {$2$};
  \node[state,fill=green!50, text=black, circle,scale=0.6, draw=black] (Nd1) [above right of=Nd2]  {$1$};
  \node[state,fill=red!50, text=black, circle,scale=0.6, draw=black] (Nd3) [below right of=Nd1] {$3$};
      \begin{pgfonlayer}{background}
    \end{pgfonlayer}
\path[-,thick] (N1) edge[thick]  (N2);
\path[-,thick] (N2) edge[thick]  (N3);
\path[-,thick] (N1) edge[thick]  (N3);
\draw[->,thick] (Nd1) to (Nd3);
\draw[->,thick] (Nd3) to (Nd2);
\draw[->,thick] (Nd2) to (Nd1);
\end{tikzpicture}
\caption{Undirected graph $\mathcal{G}_u$ and directed graph $\mathcal{G}_d$ used for consensus problems.}\label{fig_3agents}
\end{center}
\end{figure}
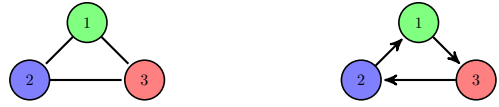
Based on the Laplacian of graph $\mathcal{G}_u$ and graph $\mathcal{G}_d$ \cite{MEbook}, the dynamics of these networks can be written as,
\begin{equation*}
\begin{split}
&\dot{\bm{x}}_u  = \bm{A}_u \bm{x}_u  =
\begin{bmatrix}
-2 &1 &1 \\
1 &-2& 1 \\
1 &1& -2
\end{bmatrix}
\bm{x}_u
,\\
&
\dot{\bm{x}}_d  = \bm{A}_d \bm{x}_d =
\begin{bmatrix}
-1 &0 &1 \\
1 &-1 &0 \\
0 &1 &-1
\end{bmatrix}
\bm{x}_d.
\end{split}
\end{equation*}
Both networks have symmetry with respect to,
\begin{equation*}
\bm{P}= \begin{bmatrix}
0 & 0 & 1 \\ 1 & 0 & 0 \\ 0 & 1& 0 
\end{bmatrix}.
\end{equation*}
The Koopman eigenfunctions and eigenvalues of the systems are obtained according to Section 4.1 in \cite{Williams_JNS_2015}.
The undirected network has a rotational eigenfunction and a pair of reflectional eigenfunctions as
\begin{equation*}
\begin{split}
&\psi_{u1} = x_{u1} +x_{u2} +x_{u3},
\\
&
 \psi_{u2}  = -1.3223  x_{u1}+  1.0954  x_{u2} +  0.2269 x_{u3},
 \\
&
\psi_{u3} =  1.0954  x_{u1} +  0.2269 x_{u2} -1.3223 x_{u3},
\end{split}
\end{equation*}
with the corresponding Koopman eigenvalues as $$\lambda_{u1} =0, \lambda_{u2} =-3, \lambda_{u3} =-3,$$ respectively.
The directed network has three rotational eigenfunctions as
\begin{equation*}
\begin{split}
&\psi_{d1} = x_{d1} +x_{d2} +x_{d3},
\\
&
 \psi_{d2}  =  x_{d1}+ e^{\mathbb{j} 2 \pi/3} x_{d2} +e^{- \mathbb{j} 2 \pi/3}x_{d3},
 \\
&
\psi_{d3} =  x_{d1}+ e^{-\mathbb{j} 2 \pi/3} x_{d2} +e^{\mathbb{j}2 \pi/3} x_{d3},
\end{split}
\end{equation*}
with the corresponding Koopman eigenvalues as $$\lambda_{d1} =0, \lambda_{d2} =-\sqrt{3} e^{- \mathbb{j} 5 \pi/6}, \lambda_{d3} =-\sqrt{3} e^{\mathbb{j} 5 \pi/6},$$ respectively.
Since the undirected network has the pair of reflectional eigenfunctions, the Koopman set includes eigenfunctions with a repeated eigenvalue with multiplicity 2.
Now, we study the observability problem of networks with respect to the measurements as 
\begin{equation*}
y_u= x_{u1}, ~~~~~ y_d= x_{d1}.
\end{equation*}
Since the undirected network has the repeated eigenvalue with multiplicity two, then the system is not observable according to Corollary \ref{Corollary_Obs_Auto_Syemmetry}.
However, the directed network has only rotational eigenfunctions, therefore the system is observable, based on Theorem \ref{Theorem_obs_original_sys}.
Now, let us consider the measurement such that it is also symmetric with respect to $\bm{P}$ as,
\begin{equation*}
\begin{split}
&y_u=x_{u1} +   x_{u2}  +x_{u3},
\\
&y_d= x_{d1}+x_{d2}+x_{d3}.
\end{split}
\end{equation*}
Therefore, both directed and undirected networks are unobservable, according to Theorem \ref{Theorem_Obs_Sys_Syemmetry}.
These results are supported using the linear algebraic conditions for the observability problem of linear systems \cite{MEbook}.
}

\textit{Example 2}.
Consider the dynamic system,
\begin{equation} \label{eq_example_1}
\begin{split}
 \dot{x}_1(t)  ~=&~~  x_1 (t) ,  \\
 \dot{x}_2(t)  ~= &~~ x_2 (t) ,\\
 \dot{x}_3(t)  ~= &~~-2 x_1^2 (t) -2 x_2^2 (t) + 4 x_3 (t), \\
 y(t) ~=&~~ x_1^2 (t) +x_2^2 (t) + x_3(t);
\end{split}
\end{equation}
see~\cite{Surana_CDC_2016}.
This system can be written in the form (\ref{eq_auto_system}) by defining $\bm{x}(t) = \left[ x_1(t), x_2(t), x_3(t)  \right]^\top$. 
The construction of the basis eigenfunctions $\Psi_b$ is inspired by \cite{Surana_CDC_2016};
in this case we let $\Psi_b =\left\lbrace \psi_1, \psi_2, \psi_3, \psi_4, \psi_5 \right\rbrace$, where, 
\begin{equation*}
\begin{split}
&\psi_1 = x_1 (t),\psi_2  = x_2 (t), \psi_3 = x_1^2 (t),  \psi_4= x_2^2 (t),
\\
&
 \psi_5 = - x_1^2 (t) - x_2^2 (t) + x_3 (t),
\end{split}
\end{equation*}
with the corresponding Koopman eigenvalues as $$\lambda_1 =1, \lambda_2 =1, \lambda_3 =2, \lambda_4 =2, \lambda_5 =4,$$ respectively. For the case of full-state observable $\bm{\Upsilon}  \left( \bm{x}(t) \right) = \bm{x}(t) $, the
Koopman modes are,
$$
\bm{v}_1 =  
\begin{bmatrix}
1 \\
0  \\
0 
\end{bmatrix},
\bm{v}_2 =  
\begin{bmatrix}
0 \\
1  \\
0 
\end{bmatrix},
\bm{v}_3 =  
\begin{bmatrix}
0 \\
0  \\
1 
\end{bmatrix},
\bm{v}_4 =  
\begin{bmatrix}
0 \\
0  \\
1 
\end{bmatrix},
\bm{v}_5 =  
\begin{bmatrix}
0 \\
0  \\
1 
\end{bmatrix}.$$
In the meantime, the system (\ref{eq_example_1}) is symmetric with respect to the permutation matrix
\begin{equation*}\label{eq_example_1_per}
\bm{P}=
\begin{bmatrix}
0 & 1  & 0 \\
1  &0  &0 \\
0  &0 &  1
\end{bmatrix}.
\end{equation*}
Theorem \ref{Theorem_Koopman_symmetry} and Corollary \ref{Corollary_Koopman_symmetry} now imply that {symmetry in the dynamics is reflected in} Koopman eigenfunctions, modes, and eigenvalues as,
\begin{equation*}
\begin{split}
&\psi_1 (\bm{x}) =\psi_2 (\bm {P} \bm{x}), \bm{v}_1= \bm{P} \bm{v}_2, \lambda_1 = \lambda_2, \\
&\psi_2 (\bm{x}) =\psi_1 (\bm {P} \bm{x}), \bm{v}_2= \bm{P} \bm{v}_1, \lambda_2 = \lambda_1, \\
&\psi_3 (\bm{x}) =\psi_4 (\bm {P} \bm{x}), \bm{v}_3= \bm{P} \bm{v}_4, \lambda_3 = \lambda_4 \\
&\psi_4 (\bm{x}) =\psi_3 (\bm {P} \bm{x}), \bm{v}_4= \bm{P} \bm{v}_3, \lambda_4 = \lambda_3 \\
&\psi_5 (\bm{x}) =\psi_5 (\bm {P} \bm{x}), \bm{v}_5= \bm{P} \bm{v}_5.
\end{split}
\end{equation*}
Furthermore, the nonlinear system (\ref{eq_example_1}) can be written in the form of the linear system,
{\begin{equation}\label{eq_example_2}
\begin{split}
 \dot{\bm{z}}(t)& =\left( \bm{ \Lambda} \otimes \bm{I} \right) \bm{z}(t)  = \left( \begin{bmatrix}
1 &0  &0  &0  &0  \\
0 &1  &0  &0  &0 \\
0 &0  &2  &0  &0 \\
0 &0  &0  &2  &0 \\
0 &0  &0  &0  &4 \\
\end{bmatrix} \otimes \bm{I} \right) \bm{z}(t),\\
\bm{y} (t) & = \left(\bm{C} \otimes \mathbbm{1}^\top \right) \bm{z} (t) = 
\left(
\begin{bmatrix}
0 & 0  & 2 & 2  & 1
\end{bmatrix} \otimes \mathbbm{1}^\top \right)\bm{z} (t).
\end{split}
\end{equation}
The measurement $h \big( \bm{x } (t) \big) = x_1^2 (t) +x_2^2 (t) +x_3 (t)$ satisfies $h \big( \bm{P} \bm{x} \big) = h \big( \bm{x} \big)$.
It thus follows that (\ref{eq_example_2}) is symmetric with respect to,
$$
 \bm{Q} \otimes \bm{I} = \begin{bmatrix}
0 &1  &0  &0  &0  \\
1 &0  &0  &0  &0 \\
0 &0  &0  &1  &0 \\
0 &0  &1  &0  &0 \\
0 &0  &0  &0  &1
\end{bmatrix}  \otimes \bm{I} .$$}
We now note that according to Theorem \ref{Theorem_Obs_Sys_Syemmetry}, the nonlinear system (\ref{eq_example_1}) is unobservable.
Figure \ref{Fig_example1_fig1} demonstrates this as two different initial conditions $\bm{x}_1(0) = [1,2,1]^\top$ and $\bm{x}_2(0) = [2,1,1]^\top$ lead to identical measurement time histories.
\begin{figure}[!ht]
\begin{center}
\includegraphics[trim={1cm 0.5cm 1cm .5cm},clip,scale=.5]{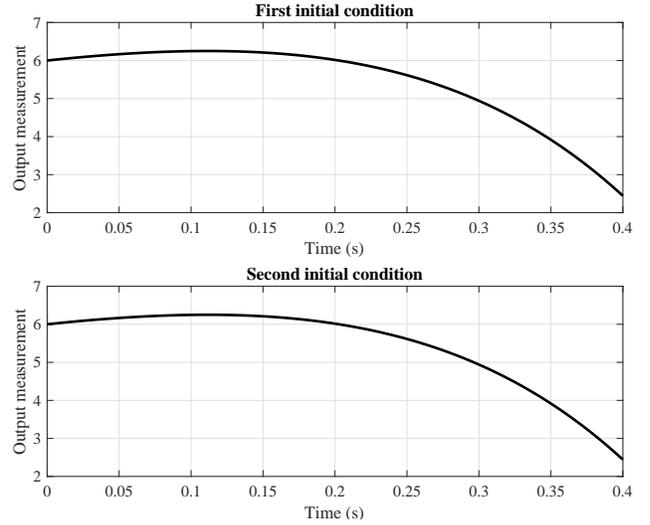}
\caption{A visual representation of the measurement time history for two different initial conditions where $\bm{x}_1(0) = \bm{P} \bm{x}_2(0)$. The two initial conditions are not distinguishable form the system measurement.}
\label{Fig_example1_fig1}
\end{center}
\end{figure}
In this case, there are two Koopman eigenvalues with multiplicity 2.
As such, at least two measurements are required to make the system observable-- this is a direct consequence of {Corollary \ref{Corollary_Obs_Auto_Syemmetry} }.
Furthermore, {Theorem \ref{Theorem_Obs_Sys_Syemmetry}} suggests that the measurements should not be symmetric with respect to $\bm{P}$.
Let us examine the system measurement of the form,
\begin{equation*}
\bm{\bar{y}} (t) = 
\begin{bmatrix}
2 x_1 (t) -  x_2^2 (t) + x_3 (t) \\
-  x_1^2 (t) + x_2 (t) + x_3 (t)
\end{bmatrix}.
\end{equation*}
Using the Koopman eigenfunctions, the system (\ref{eq_example_1}) can be expanded in the form of the linear system,
{\begin{equation}\label{eq_example_3}
\begin{split}
 \dot{\bm{z}}(t)& = \left( \bm{ \Lambda} \otimes \bm{I} \right) \bm{z}(t),\\
\bm{\bar{y}}  (t) & = \left( \bm{\bar{C}} \otimes \mathbbm{1}^\top \right)   \bm{z} (t) = \left( \begin{bmatrix}
2 & 0 & 1 & 0 & 1 \\
0 & 1 & 0 & 1 & 1 
\end{bmatrix}  \otimes \mathbbm{1}^\top \right) \bm{z} (t).
\end{split}
\end{equation}
In this case, since $\left( \bm{\bar{C}}  \otimes \mathbbm{1}^\top \right) \neq \left( \bm{Q}  \otimes \bm{I} \right) \left( \bm{\bar{C}}  \otimes \mathbbm{1}^\top \right)$, the linear system (\ref{eq_example_3}) is not symmetric with respect to $\bm{Q} \otimes \bm{I}$}.
In fact, (\ref{eq_example_3}) and the nonlinear system (\ref{eq_example_1}) are both observable as conditions (\ref{eq_condition_obser}) and (\ref{eq_condition_obser_origi}) are satisfied,  and Lemma \ref{Lemma_observability} and Theorem \ref{Theorem_obs_original_sys} become applicable.

Although, the initial conditions $\bm{x}_1(0) = [1,2,1]^\top$ and $\bm{x}_2(0) = [2,1,1]^\top$ are not distinguishable from the measurement $\bm{y}(t)$, Figure \ref{Fig_example1_fig2} depicts how the 		``non-symmetric" measurement $\bm{\bar{y}}(t)$ can distinguish the two distinct initial conditions.
\begin{figure}[!ht]
\begin{center}
\includegraphics[trim={1cm 0.5cm 1cm .5cm},clip,scale=.5]{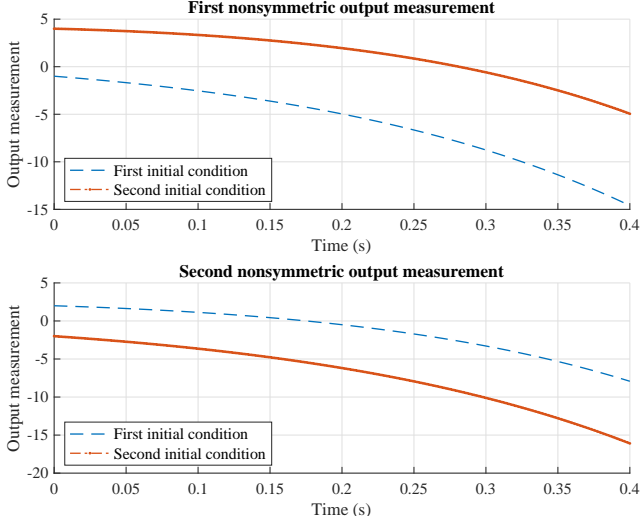}
\caption{Two distinct initial conditions for which $\bm{x}_1(0) = \bm{P} \bm{x}_2(0)$ are distinguishable from non-symmetric measurements.}
\label{Fig_example1_fig2}
\end{center}
\end{figure}

\textit{Example 3}.
Consider a ring of eight reactively coupled nanoelectromechanical oscillators \cite{Emenheiser_Chaos_2016,matheny2019}, depicted in Figure \ref{fig_example_network}, with the local dynamics governed as,
\begin{equation}\label{eq_example_rings}
\frac{d x_i}{dt}=-\frac{1}{2} x_i + \mathbb{j} | x_i |^2 x_i+\frac{x_i}{2 |x_i|}+ \mathbb{j} \frac{\beta}{2} \big( x_{i+1} - 2 x_i + x_{i-1}\big),
\end{equation}
where $x_i \in \mathbb{C}$ denotes the amplitude and phase of the $i$-th oscillator, $x_0=x_8$, and $x_9=x_1$, for $i \in 1,\ldots, 8$.


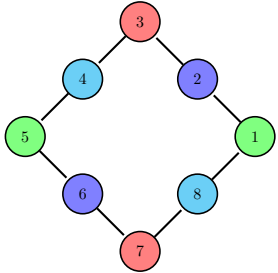
\begin{figure}[!ht]
\begin{center}
\begin{tikzpicture}[->,>=stealth',shorten >=1pt,auto,node distance=1.8cm,
                    semithick]
  \node[state,fill=green!50, text=black, circle,scale=0.6, draw=black] (N1)                                   {$1$};
  \node[state,fill=blue!50, text=black, circle,scale=0.6, draw=black] (N2) [above left of=N1] {$2$};
  \node[state,fill=red!50, text=black, circle,scale=0.6, draw=black] (N3) [above left of=N2] {$3$};
  \node[state,fill=cyan!50, text=black, circle,scale=0.6, draw=black] (N4) [below left of=N3] {$4$};
  \node[state,fill=green!50, text=black, circle,scale=0.6, draw=black] (N5) [below left of=N4] {$5$};
  \node[state,fill=blue!50, text=black, circle,scale=0.6, draw=black](N6) [below right of=N5] {$6$};  
  \node[state,fill=red!50, text=black, circle,scale=0.6, draw=black](N7) [below right of=N6] {$7$}; 
  \node[state,fill=cyan!50, text=black, circle,scale=0.6, draw=black](N8) [above right of=N7] {$8$}; 
      \begin{pgfonlayer}{background}
    \end{pgfonlayer}
\path[-] (N1) edge[thick]  (N2);
\path[-] (N2) edge[thick]  (N3);
\path[-] (N3) edge[thick]  (N4);
\path[-] (N4) edge[thick]  (N5);
\path[-] (N5) edge[thick]  (N6);
\path[-] (N6) edge[thick]  (N7);
\path[-] (N7) edge[thick]  (N8);
\path[-] (N8) edge[thick]  (N1);
\end{tikzpicture}
\caption{A visual representation of the ring of NEMs oscillators. The presence of network symmetry is depicted using same colored nodes for encoding the underlying automorphism.}\label{fig_example_network}
\end{center}
\end{figure}
The complex-valued weighted nonlinear representation of this network (\ref{eq_example_rings}) can be decomposed in terms of its amplitude and phase components as,
\begin{equation}\label{eq_example2_amp_pha}
\begin{split}
\frac{d a_i}{dt} = &  -\frac{a_i-1}{2} - \frac{\beta}{2} \big( a_{i+1} \text{sin} (\phi_{i+1}-\phi_i)  \\ &~~~~~~~~~~~~~~~~~~~~~~~~~+   a_{i-1} \text{sin} (\phi_{i-1}-\phi_i) \big),\\
\frac{d \phi_i}{dt}  =&   \alpha a_i^2+ \frac{\beta}{2 a_i} \big( a_{i+1} \text{cos} (\phi_{i+1}-\phi_i)  \\ &~~~~~~~~~~~~~~~~~~~~+   a_{i-1} \text{cos} (\phi_{i+1}-\phi_i) - 2\big),
\end{split}
\end{equation}
where $x_i=a_i e^{\mathbb{j} \phi_i}$ such that $a_i \in \mathbb{R}$ and $\phi_i \in \mathbb{R}$ are, respectively, the amplitude and phase of the $i$-th oscillator, for $i \in 1,\ldots, 8$.
Let us define the measurement as $$h \big( x_1,\ldots,x_8 \big) = \sum_{i=1}^8 \text{cos} ( \phi_{i}-\phi_{i+1} ).$$
Hence,  the structure of the network and the output measurement $h$ have a reflectional symmetry with respect to,
$$
\bm{P}  = \begin{bmatrix}
0 &0  &0  &0  &1 &0  &0  &0  \\
0 &0  &0  &0  &0 &1  &0  &0  \\
0 &0  &0  &0  &0 &0  &1  &0  \\
0 &0  &0  &0  &0 &0  &0  &1  \\
1 &0  &0  &0  &0 &0  &0  &0  \\
0 &1 &0  &0  &0 &0  &0  &0  \\
0 &0  &1  &0  &0 &0  &0  &0  \\
0 &0  &0  &1  &0 &0  &0  &0
\end{bmatrix}.$$

Now we examine the simulation results for two different initial conditions, $\bm{x}_1(0) = [x_1(0),\ldots,x_8(0)]^\top$ and $\bm{x}_2(0) = \bm{P} \bm{x}_1(0)$, for $\alpha = 1$ and $\beta = 0.1$.
Figures \ref{Fig_example2_amplitudes} and \ref{Fig_example2_phases} demonstrate the amplitude and phase trajectories in a ring of eight oscillators with the coupling $\beta = 0.1$ and nonlinearity $\alpha = 1$.
The amplitude and phase trajectories of the $1$st, $2$nd, $3$rd, and $4$th  oscillators for the first initial condition are identical to those of $5$th, $6$th, $7$th, and $8$th oscillators, respectively.
\begin{figure}[!ht]
\begin{center}
\includegraphics[trim={1cm 0.5cm 1cm .5cm},clip,scale=.5]{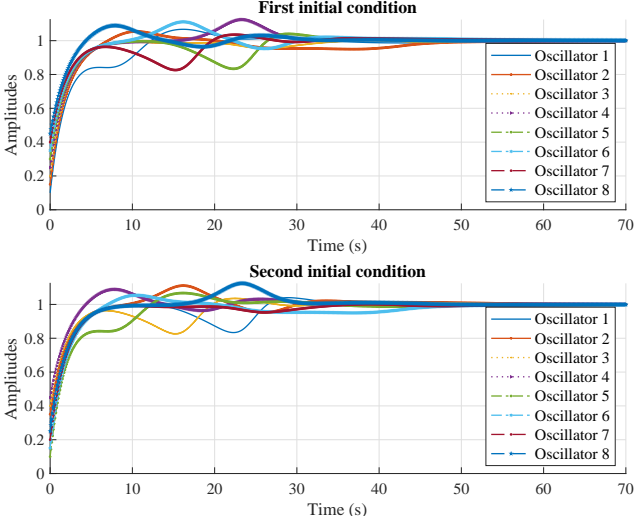}
\caption{A visual representation of the amplitude trajectories of NEMs for two reflectional initial conditions.}
\label{Fig_example2_amplitudes}
\end{center}
\end{figure}
\begin{figure}[!ht]
\begin{center}
\includegraphics[trim={1cm 0.5cm 1cm .5cm},clip,scale=.5]{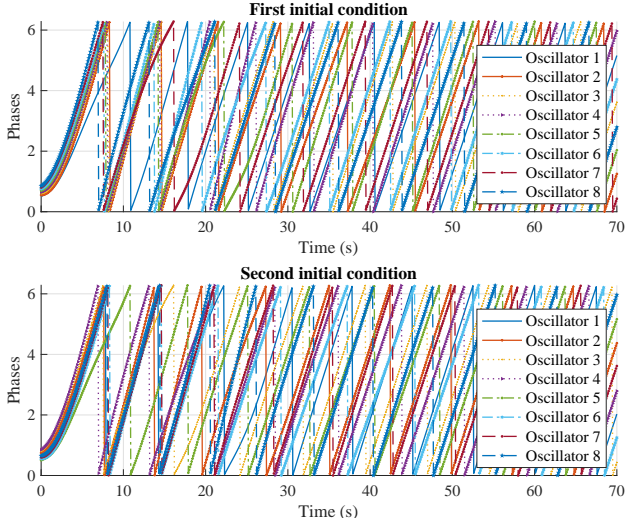}
\caption{A visual representation of the phase trajectories of NEMs for two reflectional initial conditions.}
\label{Fig_example2_phases}
\end{center}
\end{figure}
We note that the NEMs network (\ref{eq_example2_amp_pha}) is not ``projectively" symmetric with respect to $\bm{P}$.
As such, Theorem \ref{Theorem_Obs_Sys_Syemmetry}
now implies that the NEM network in a ring topology, shown in Figure \ref{fig_example_network}, is unobservable.
Figure \ref{Fig_example2_2outputs} shows the (indistinguishable) measurements of this system for two distinct initial conditions, $\bm{x}_1(0) = [x_1(0),\ldots,x_8(0)]^\top$ and $\bm{x}_2(0) = \bm{P} \bm{x}_1(0)$, 
respectively.
\begin{figure}[!ht]
\begin{center}
\includegraphics[trim={1cm 0.5cm 1cm .5cm},clip,scale=.5]{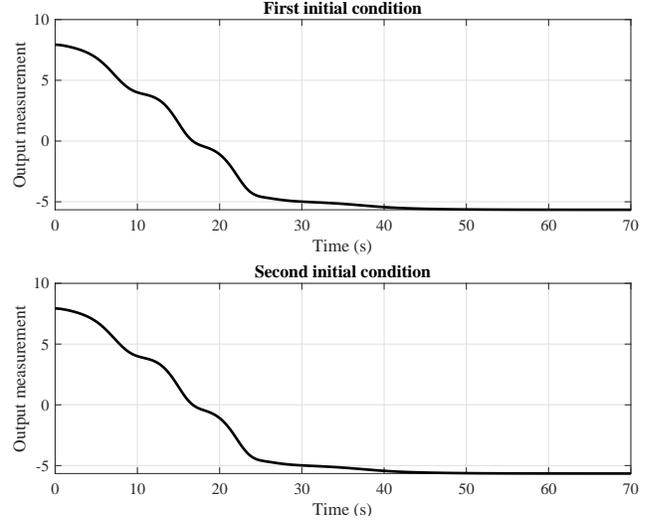}
\caption{A visual representation of the output measurements for two different initial conditions $\bm{x}_1(0) = [x_1(0),\ldots,x_8(0)]^\top$ and $\bm{x}_2(0) = \bm{P} \bm{x}_1(0)$. These sets of initial conditions are not distinguishable form the corresponding measurements.}
\label{Fig_example2_2outputs}
\end{center}
\end{figure}


\section{Concluding Remarks} \label{Section_Conclusion}
This paper presents an approach for examining nonlinear observability 
in a Koopman operator-theoretic framework, with less emphasis on geometrical and algebraic approaches typically adopted to examine this problem.
This is achieved by transforming the nonlinear system into an infinite-dimensional linear system based on independent Koopman eigenfunctions that facilitates determining the observability of the system via
the spectral properties of the corresponding infinite dimensional linear system.
These spectral properties are examined in terms of the rank of a finite-dimensional matrix.
{Further, we examined how {discrete} symmetries in the dynamics are reflected in the spectral
properties of the corresponding Koopman operator.
In particular, it is shown that such symmetries have implications in terms of
symmetries in the Koopman eigenspace as well as the presence of repeated eigenvalues.
%
These observations in turn enabled use to spectral methods for identifying the implications of
symmetry for nonlinear unobservability.

Future directions for this work include using the Koopman operator framework for addressing controllability of nonlinear systems. {It is also of interest} to design more efficient and accurate numerical algorithms for computing Koopman properties of symmetric dynamical systems; see~\cite{Salova_APS_2018}.
\section*{Acknowledgments}
The authors thank Mathias Hudoba de Badyn for his helpful comments and MATLAB code for simulating a network of nanoelectromechanical systems.
Discussions with Raissa D'Souza, James Crutchfield, Anastasyia Salova, and Jeff Emenheiser on the Koopman operator and symmetries are much appreciated. The example pertaining to the ring of NEMS oscillators is inspired by the work of Matt Matheny, Warren Fon, and Micheal Roukes, who have lead the design, fabrication, and testing of this network, exhibiting intricate ``exotic" states; see~\cite{matheny2019}. The authors also thank the reviewers and the Associate Editor for their helpful suggestions. This research has been supported by the U.S. Army Research
Laboratory and the U. S. Army Research Office under award W911NF-13-1-0340.
\balance 
\bibliographystyle{plain}
\bibliography{bibliography}
\end{document}